\documentclass[twocolumn, resetfootnote]{aastex7}

\expandafter\def\csname editcolor1\endcsname{magenta}
\expandafter\def\csname editcolor2\endcsname{blue}  
\expandafter\def\csname editcolor3\endcsname{violet} 

\usepackage{newtxtext,newtxmath}
\usepackage[T1]{fontenc}
\usepackage{graphicx}	
\usepackage{amsmath}	
\usepackage{float}
\usepackage{bookmark} 
\bookmarksetup{numbered, open,}
\usepackage{enumitem}
\setlist[enumerate]{itemsep=0mm}
\usepackage{hyperref}
\usepackage{subfigure}
\usepackage{xspace}
\usepackage{xcolor} 
\usepackage{soul} 
\usepackage{multirow}

\DeclareRobustCommand{\okina}{%
  \raisebox{\dimexpr\fontcharht\font`A-\height}{%
    \scalebox{0.8}{`}%
  }%
}

\newcommand{\Oumuamua}{\okina Oumuamua\xspace}

\newcommand{\FeI}{Fe~{\sc i}}

\newcommand{\NiI}{Ni~{\sc i}}

\begin{document}

\title{Contemporaneous Optical and Near-Infrared Observations of the Interstellar Comet 3I/ATLAS Pre- and Post-Perihelion}

\correspondingauthor{Kyle Medler}
\email{kmedler@hawaii.edu}

\author[0000-0001-7186-105X]{Kyle~Medler}
\affiliation{Institute for Astronomy, University of Hawai\okina i, 2680 Woodlawn Drive, Honolulu, HI 96822, USA}
\email{kmedler@hawaii.edu}

\author[0000-0003-3953-9532]{Willem~B.~Hoogendam}
\altaffiliation{NSF Graduate Research Fellow}
\affiliation{Institute for Astronomy, University of Hawai\okina i, 2680 Woodlawn Drive, Honolulu, HI 96822, USA}
\email{willemh@hawaii.edu}

\author[0000-0002-5221-7557]{Christopher~Ashall}
\affiliation{Institute for Astronomy, University of Hawai\okina i, 2680 Woodlawn Drive, Honolulu, HI 96822, USA}
\email{cashall@hawaii.edu}

\author[0000-0002-5033-9593]{Bin~Yang}
\affiliation{Instituto de Estudios Astrof\'isicos, Facultad de Ingenier\'ia y Ciencias, Universidad Diego Portales, Santiago, Chile}
\email{bin.yang@mail.udp.cl} 

\author[0000-0001-5559-2179]{James~J.~Wray}
\affiliation{School of Earth and Atmospheric Sciences, Georgia Institute of Technology, 311 Ferst Drive, Atlanta, GA 30332, USA}
\affiliation{Institute for Astronomy, University of Hawai\okina i, 2680 Woodlawn Drive, Honolulu, HI 96822, USA}
\email{jwray@gatech.edu}

\author[0000-0003-4631-1149]{Benjamin~J.~Shappee}
\affiliation{Institute for Astronomy, University of Hawai\okina i, 2680 Woodlawn Drive, Honolulu, HI 96822, USA}
\email{shappee@hawaii.edu}

\author[0000-0002-2058-5670]{Karen~J.~Meech}
\affiliation{Institute for Astronomy, University of Hawai\okina i, 2680 Woodlawn Drive, Honolulu, HI 96822, USA}
\email{meech@hawaii.edu}  

\author[0000-0002-2471-8442]{Michael~A.~Tucker}
\affiliation{Center for Cosmology and Astroparticle Physics, The Ohio State University, Columbus, OH, USA}
\affiliation{Department of Astronomy, The Ohio State University, Columbus, OH, USA}
\email{tucker.957@osu.edu}

\author[0000-0002-4449-9152]{Katie~Auchettl}
\affiliation{School of Physics, The University of Melbourne, Parkville, VIC, Australia}
\affiliation{Department of Astronomy and Astrophysics, University of California, Santa Cruz, CA, USA}
\email{katie.auchettl@unimelb.edu.au}

\author[0000-0002-2164-859X]{Dhvanil~D.~Desai}
\affiliation{Institute for Astronomy, University of Hawai\okina i, 2680 Woodlawn Drive, Honolulu, HI 96822, USA}
\email{dddesai@hawaii.edu}

\author[0000-0001-9668-2920]{Jason~T.~Hinkle}
\altaffiliation{NHFP Einstein Fellow}
\affiliation{Department of Astronomy, University of Illinois Urbana-Champaign, 1002 West Green Street, Urbana, IL 61801, USA}
\affiliation{NSF-Simons AI Institute for the Sky (SkAI), 172 E. Chestnut St., Chicago, IL 60611, USA}
\affiliation{Institute for Astronomy, University of Hawai\okina i, 2680 Woodlawn Drive, Honolulu, HI 96822, USA}
\email{jhinkle6@hawaii.edu}

\author[orcid=0000-0002-8732-6980]{Andrew~M.~Hoffman}
\affiliation{Institute for Astronomy, University of Hawai\okina i, 2680 Woodlawn Drive, Honolulu, HI 96822, USA}
\email{amho@hawaii.edu}  

\author[0000-0003-1059-9603]{Mark~E.~Huber}
\affiliation{Institute for Astronomy, University of Hawai\okina i, 2680 Woodlawn Drive, Honolulu, HI 96822, USA}
\email{mehuber7@hawaii.edu}

\author[0000-0002-6230-0151]{David~O.~Jones}
\affiliation{Institute for Astronomy, University of Hawai\okina i, 640 N. A\okina ohoku Pl., Hilo, HI 96720, USA}
\email{dojones@hawaii.edu}

\author[0000-0003-4936-4959]{Ruining~Zhao}
\affiliation{National Astronomical Observatories, Chinese Academy of Sciences, Beijing 100101, China}
\email{rnzhao@nao.cas.cn}

\begin{abstract}
Interstellar objects provide a unique view into the formation of other star systems. Here we present spectroscopic observations of the recently discovered interstellar object 3I/ATLAS between a heliocentric distance of $3.7$ to $1.8$~au on either side of its travels through perihelion. We obtained several observations with the Keck-I/LRIS, Keck-II/NIRES, Gemini/GMOS, and UH88/SNIFS spectrographs, covering a wavelength range of $0.3 - 2.5$~\micron. We report the continued emission of both Ni and CN, along with post-perihelion detections of Fe and a weak detection of $\mathrm{C_3}$. We determine the spectral slope across optical and NIR wavelengths and find a positive spectral slope in the optical, with values ranging from $\sim 21 - 27\%$ in the blue regions ($0.4 - 0.55$~\micron) to $\sim 6 - 10\%$ in the red ($0.65 - 0.9$~\micron) regions. In contrast, the NIR showed a negative spectral slope of $\sim -0.9 \%$ between $0.9 - 1.5$~\micron\ and $\sim -2.3\%$ between $1.9 - 2.5$\micron. 3I/ATLAS shows a clear turnover in its spectral shape at $\sim 1.1$~\micron, corresponding to scattered light from the dusty coma. Finally, in the NIR, we do not find an increase in the depth of the water features identified in an earlier NIR observation of 3I/ATLAS. Our observations of 3I/ATLAS in the NIR show a similar shape to the NIR spectrum of 2I/Borisov as it approached perihelion.
\end{abstract}



\section{Introduction}\label{sec:intro}

Remnant solar system planetesimals (i.e., comets, asteroids, meteors, etc.) serve as probes of the original formation conditions and composition of the Solar System (e.g., \citealp{Bergin2024}
). Interstellar objects (ISOs) are extrasolar planetesimals, allowing for the analysis of other solar systems. What makes the study of ISOs so critical is their potential to record the properties of their original solar system, e.g., providing information on chemical abundances \citep[e.g.,][]{Jewitt2023ARAA, Fitzsimmons2024}. While it is expected that several ISOs traverse through the solar system at any given time \citep{Do2018}, only three have been discovered to date: 1I/\Oumuamua \citep{Meech2017}, 2I/Borisov \citep{borisov_2I_cbet}, and, most recently, 3I/ATLAS. 3I/ATLAS was discovered 01 Jul 2025 \citep{Denneau2025} by the Asteroid Terrestrial-impact Last Alert System (ATLAS; \citealp{Tonry2018a, Tonry2025}). 
This triggered a global campaign to obtain additional observations \citep[e.g.,][]{Belyakov2025, Yang2025, Cordiner_2025, Tonry2025, Hutsemekers25, Hoogendam25_SNIFS}.

This small sample of observed interstellar objects has shown a diverse range of cometary properties. 
The first ISO 1I/\Oumuamua\ showed no cometary activity in the form of a dust coma but did exhibit non-gravitational acceleration, similar to the “dark comets” \citep{Seligman2024}.
In contrast, the second ISO, 2I/Borisov, was significantly more active, with high production rates of gasses including an unusually high CO abundance (relative to $\mathrm{H_2O}$), $>3$ times that of inner solar system comets \citep[e.g.,][]{Jewitt2023ARAA, Fitzsimmons2024}, as well as a dusty coma \citep{Fitzsimmons:2019, Guzik_2020, Jewitt2019b, Opitom:2019-borisov}.
 
Compared to the other ISOs, 3I/ATLAS is similar to 2I/Borisov, possessing a large dust coma and high gas production as it approached the sun \citep{Seligman2025, Jewitt2025, Cordiner_2025, delaFuenteMarcos2025, Chandler2025, Lisse_2025}. Cometary activity has been observed in various wavelength regimes, including water and OH in the UV \citep{Xing2025}, Ni, Fe, $\mathrm{C_2}$, $\mathrm{C_3}$, CN, pre-perihelion \citep{Rahatgaonkar_2025, Hutsemekers25, Hoogendam25_SNIFS, SalazarManzano2025, Hoogendam25_KCWI}, and post-perihelion CH \citep{Hoogendam26_KCWI} in the optical, potential water ice in the near-infrared \citep[NIR;][see also \citealt{Kareta2025}]{Yang2025}, $\mathrm{CO_2}$, $\mathrm{CO}$, $\mathrm{H_2O}$, and $\mathrm{CH_4}$ in the mid-infrared \citep{Cordiner_2025, Lisse_2025, Lisse2025, Belyakov26}, and HCN and $\mathrm{CH_3OH}$ \citep{Hinkle25_JCMT, Coulson25, Roth25} in the sub-mm. Understanding the composition of 3I/ATLAS will provide insights about its formation environment and, therefore, the small bodies and planet-forming environment around another star.

Of particular interest is the water content of 3I/ATLAS. The most abundant solid in proto-planetary discs is water, and it significantly influences planetary formation processes \citep{Bitsch16, Muller21} and likely contributed to giant planet cores \citep{Ciesla14}. Furthermore, water is a key ingredient for life in our own Solar System. However, the origin of water on Earth remains debated \citep[for a review, see, e.g.,][]{Mottl07}. One model is that comets delivered water to Earth during the heavy bombardment \citep[e.g.,][]{Chyba_1990a, Meech2020}. 
This makes understanding the water activity and water content of interstellar objects like 3I/ATLAS important---do other stellar systems contain similar small bodies capable of delivering water to potentially habitable rocky planets? 

Both 2I/Borisov and 3I/ATLAS show substantial water production \citep[e.g.,][]{Yang2020, Xing2020, Yang2025, Xing2025, Cordiner_2025, Combi25}, and dynamical models suggest 3I/ATLAS should be water-rich \citep{Hopkins2025b}. In particular, 3I/ATLAS exhibited a larger inferred water active fraction (i.e., the fraction of the nucleus surface undergoing sublimation) than is typical for many Solar System comets. \citep{Xing2025, Tan26}. However, both comets showed atypical volatile compositions: 2I/Borisov had a high CO/$\mathrm{H_2O}$ ratio \citep{Cordiner_2020} and similarly 3I/ATLAS has a high $\mathrm{CO}_2$/$\mathrm{H_2O}$ ratio \citep{Cordiner_2025, Belyakov26}. Given the importance of understanding the $\mathrm{H_2O}$ content in interstellar objects, it is natural to turn to the NIR regime that hosts several signatures of water. In particular, the $\mathrm{H_2O}$ 1.5~$\mu$m absorption and 2.0~$\mu$m band are readily observed from ground-based facilities and provide direct insights into the water composition of comets. 
To that end, we present optical to NIR panchromatic observations of 3I/ATLAS obtained by the Hawai\okina i Infrared Supernova Study \citep[HISS;][]{Medler25_HISS}. 

\section{Data}\label{sec:data}
We observed 3I/ATLAS using the 10.0\,m Keck\,I and Keck\,II telescopes, the 8.1\,m Frederick C.\ Gillett Gemini Telescope, and the University of Hawai\okina i 88 inch Telescope on Maunakea. The data reduction procedures for each instrument are described below, and an observing log is provided in Table~\ref{tab:Obs_log}. The optical and NIR spectra of 3I/ATLAS are shown in Figure~\ref{fig:pan_spec}.

\begin{deluxetable*}{ccccccc}
    \tablenum{1}
    \tablecaption{Log of spectroscopic observations of 3I/ATLAS. $r_h$ is the heliocentric distance and $\Delta$ is the geocentric distance.}\label{tab:Obs_log}
    \tablewidth{0pt}
    \tablehead{ \colhead{Instrument} & \colhead{UT Date} & \colhead{$r_h$} & \colhead{$\Delta$} & \colhead{Phase Angle} &\colhead{Solar Analogues} & \colhead{True Anomaly} \\
    \colhead{} & \colhead{} & \colhead{[au]} & \colhead{[au]} & \colhead{[Degrees]} & \colhead{} & \colhead{[Degrees]}
    }
    \startdata
    LRIS  & 2025-07-26 & 3.68 & 2.90 & 11.47 & HD~146396/HD~165290 & 285.4 \\ 
    GMOS  & 2025-07-27 & 3.64 & 2.88 & 11.91 & HD~153631/HD~165290 & 285.7 \\
    GMOS  & 2025-07-31 & 3.51 & 2.82 & 13.54 & HD~153631/HD~165290 & 286.6\\
    LRIS  & 2025-08-27 & 2.66 & 2.59 & 22.13 & HD~146396/HD~165290 & 295.5\\ 
    NIRES & 2025-08-28 & 2.63 & 2.59 & 22.31 & HD~165290 & 295.9\\ 
    SNIFS & 2025-11-30 & 1.79 & 1.92 & 30.60 & BD~+01~2690 & 44.2 \\
    NIRES & 2025-11-30 & 1.79 & 1.92 & 30.60 & BD~+01~2690 & 44.2 \\
    \enddata
\end{deluxetable*}

\subsection{Optical Observations}
Four optical spectra of 3I/ATLAS were obtained over three months, from 26$^\mathrm{th}$ Jul to 30$^\mathrm{th}$ Nov 2025, as it approached and passed perihelion (Oct. 29, 2025 at 1.356 au). The reduction process for each spectrum is detailed below.

\subsubsection{LRIS Reductions}
Two optical spectra of 3I/ATLAS covering $\sim0.3 - 1.0$~\micron\ were obtained on 26.29 Jul 2025 and 27.24 Aug 2025 using the Low-Resolution Imaging Spectrometer \citep[LRIS;][]{Oke:1995, McCarthy:1998} mounted on the 10.0-m Keck-I telescope. Observations were taken using the $400/3400$ grism and $400/8500$ grating. In addition to 3I/ATLAS, observations of two solar analogs were obtained for both epochs, HD~159662 and HD~157378 were obtained for the July observation and HD~146396 and HD~165290 were observed for the August. Finally, a flux calibration star, BD+33~2642, was obtained on the 27$^\mathrm{th}$ of August at a similar airmass to 3I/ATLAS. Due to the limited time available on the 26$^\mathrm{th}$ of July, it was not possible to obtain a flux calibration standard. Instead, the same flux-calibration star from August was used to calibrate both epochs. Comparing the flux of both the comet observations and the observations of the solar analogues from the two epochs, we find agreement between the spectra of the two epochs.

The LRIS observations were reduced using the reduction code \texttt{PypeIt} \citep{ Prochaska_2020zndo, Prochaska_2020} following the method described in \citet{Medler25_HISS}. Once the blue and red arms of the LRIS observations were reduced, the arms were stitched together by aligning the high signal-to-noise (S/N) overlapping regions and cutting off the fringe of each arm. Then the combined spectrum was sigma-clipped, while masking regions where emission features from cometry activity lay, to remove low S/N regions at the edges of the grating/grism and to eliminate noise spikes in the data. 

\subsubsection{GMOS Reduction}

Another optical spectrum was obtained by the Gemini-North Multi-Object Spectrograph \citep[GMOS;][]{Hook_2004} mounted on the 8.1~m Frederick C. Gillette Gemini-North telescope. The observation of 3I/ATLAS was split between two nights, with the first observation obtained on 27.41 Jul using the B480 grating and the second observation occurring on 31.33 Jul with the R150 grating. Alongside observations of 3I/ATLAS, two solar analogues, HD~153631 and HD~165290, were also observed both nights. We flux calibrated the spectrum using a flux-standard star observed at the start of the semester, following standard Gemini operations. Observations of 3I/ATLAS and the two solar analogues were reduced using the automatic reduction pipeline DRAGONS \citep{Labrie_DRAGONS_2023, Labrie_DRAGONS_2023zndo} version 4.1.0. 

\begin{figure*}
    \centering
    \includegraphics[width=\linewidth, trim={2cm, 0cm, 3cm, 0.4cm}, clip=True]{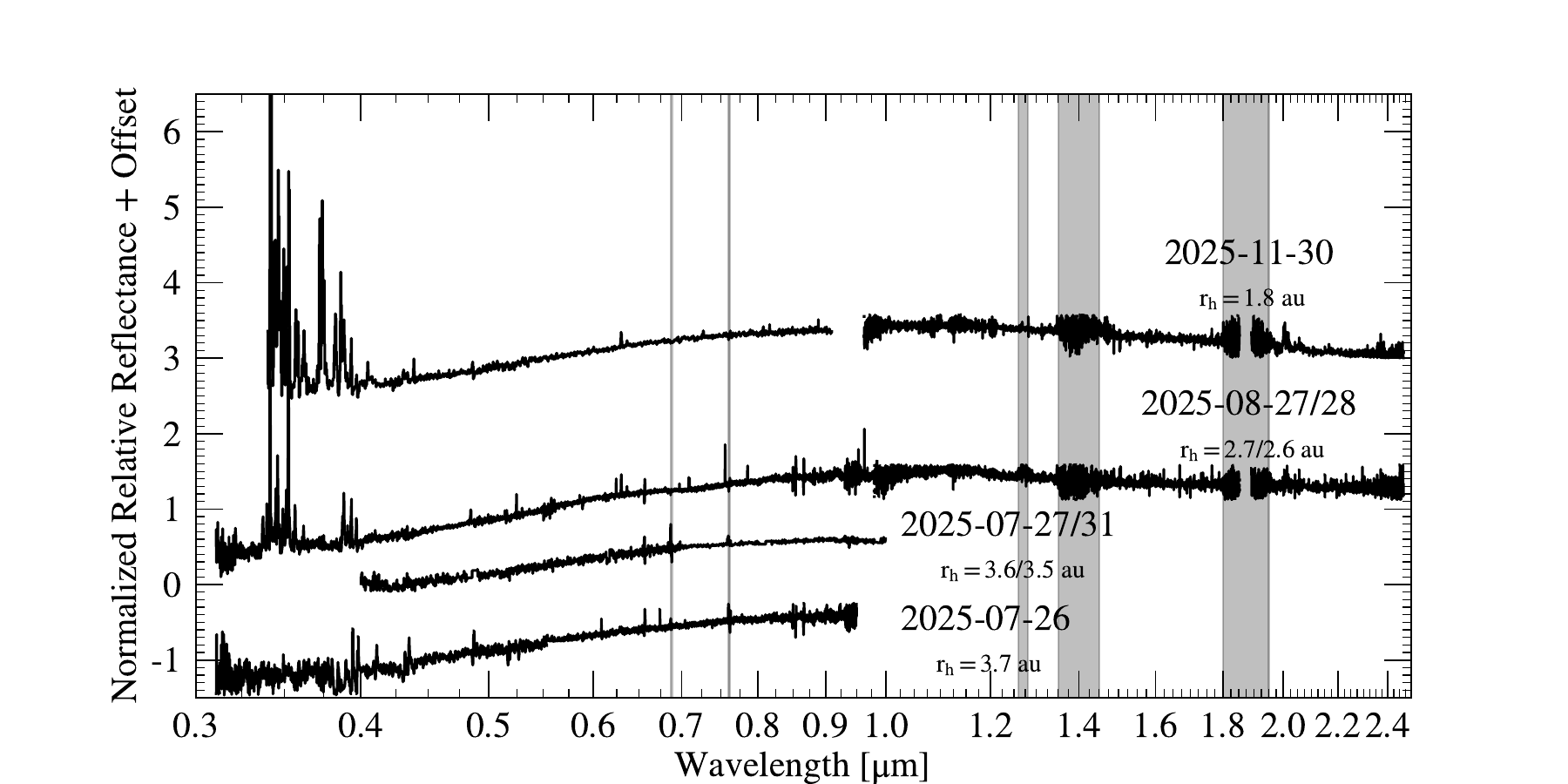}
    \caption{
    Panchromatic reflectance spectra of 3I/ATLAS. All epochs show a clear rising continuum in the optical, peaking around $1.0$~\micron, indicative of a dusty coma. Between the second and third epochs, strong Ni and CN emission lines appear blueward of $\sim 0.4$~\micron, with Fe emission lines emerging as 3I/ATLAS made its perihelion passage. All spectra are normalized to reflectance at $0.5$~\micron, and the telluric regions are denoted by the vertical shaded regions. A zoomed in view on the cometary emission region $0.33 - 0.5$~\micron\ is shown in Figure~\ref{fig:emission_features}.}
    \label{fig:pan_spec}
\end{figure*}

\subsubsection{SNIFS Reduction}

An additional optical spectrum was obtained on 30.61 Nov 2025 by the Supernova Integral Field Spectrograph \citep[SNIFS;][]{Lantz2004} mounted on the University of Hawai\okina i 88 inch Telescope. The SNIFS instrument is split into two channels, a blue channel which covers 3400--5100~\AA\ range, and a red channel which covers 5100--10000~\AA\ \citep{Aldering_2007}. The 30.61 Nov 2025 SNIFS observation was reduced using the standard SCAT pipeline as described in \citet{Tucker2022}. 

\subsection{Infrared Observations}
Several NIR spectra were obtained to closely coincide with the optical observations, enabling us to construct, for the first time, a panchromatic time series of spectra of an interstellar object spanning \(\sim 0.33\text{--}2.5\)~\micron\ both before and after perihelion.

\subsubsection{NIRES Reductions}

Two NIR spectra, covering $\sim0.7-2.5$~\micron, were obtained by the Hawai\okina i Infrared Supernova Study (HISS; \citealp{Medler25_HISS})\footnote{HISS normally observes supernovae and other astrophysical transients \citep[e.g.,][]{Hoogendam25_epr, Hoogendam25_pxl, Medler25_HISS}, but the excitement around 3I/ATLAS was contagious!} using the Near-Infrared Echellette Spectrometer \citep[NIRES;][]{Wilson_2004} on the 10.0-m Keck-II telescope on 28.22 Aug 2025 and 30.66 Nov 2025. The NIRES observations were reduced with the \texttt{Spextool} software package \citep{Vacca03, Cushing04} modified for NIRES reductions. Telluric corrections were performed using solar analogue stars in place of the typical A0V star \citep[using the procedures implemented in \texttt{xtellcor\_general}, see][for more information]{Vacca03} taken the same night with a similar airmass at the time of observations to that of 3I/ATLAS. The spectral trace in each order was extracted from the NIRES images using an aperture size of $\sim 1.0$\arcsec. The individual traces were co-added, flux-calibrated, and merged into a single spectrum.

\section{Results}\label{sec:results}

\begin{figure*}
    \centering
    \includegraphics[width=\linewidth, trim={1.5cm, 0.5cm, 3cm, 1cm}, clip=True]{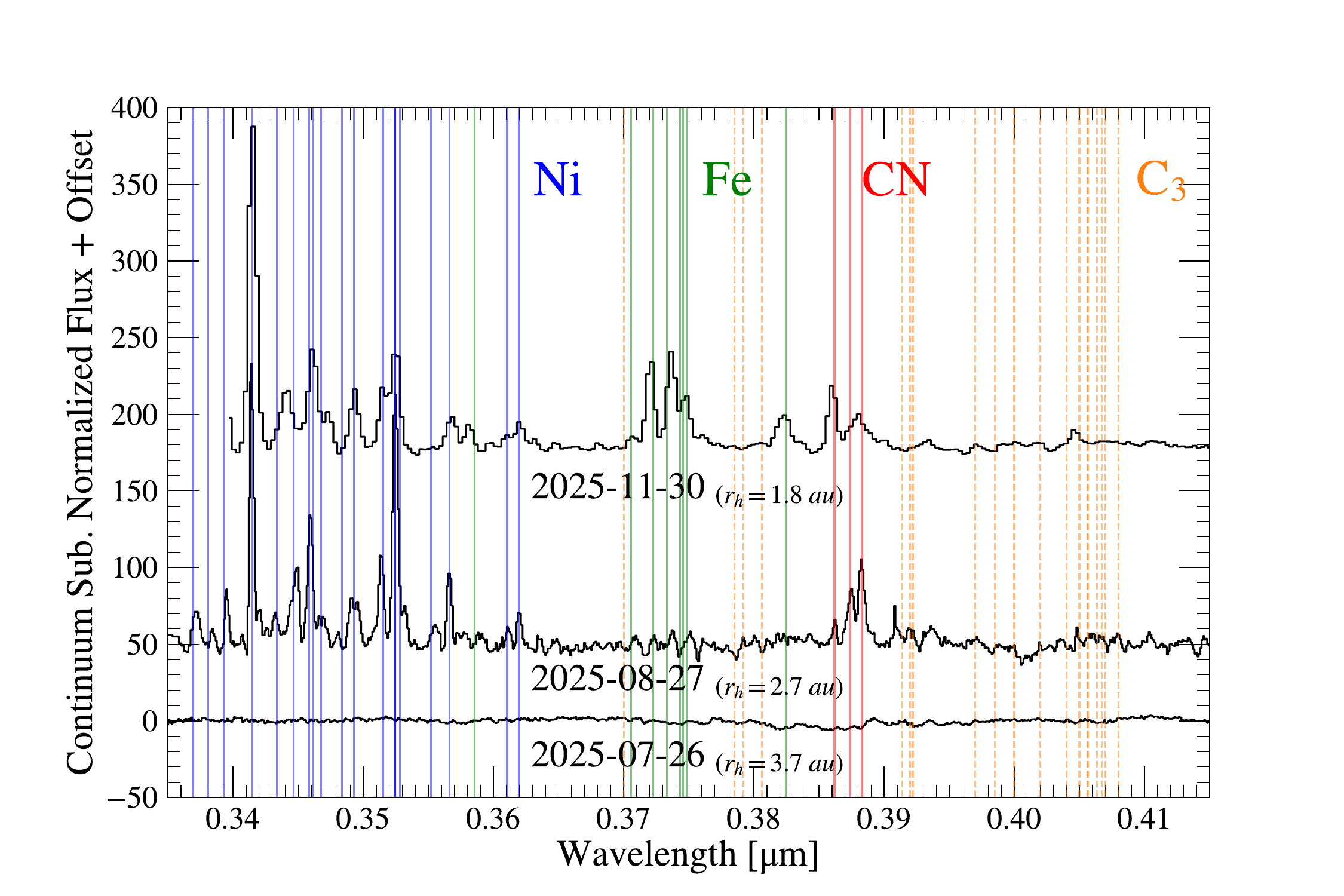}
    \caption{ 
    Emission lines associated with Ni, Fe, CN, and $\mathrm{C_3}$ that emerge in the LRIS and SNIFS spectra of 3I/ATLAS once it had moved within $\mathrm{r_h} \sim 3$~au. The Fe and $\mathrm{C_3}$ emission lines evolve as 3I/ATLAS approached and past perihelion suggesting a change in the emitting surface. This may have been caused by the Fe-rich material being buried deeper in the surface than the Ni-rich compounds.}
    \label{fig:emission_features}
\end{figure*}

The panchromatic observations of 3I/ATLAS, presented in Figure~\ref{fig:pan_spec}, show several distinct spectral features. These features range from a strong red optical spectral slope, which is observed in all spectra, to a transition into a negative spectral slope across the NIR. Additionally, strong emission lines from Ni, Fe, $\mathrm{C_3}$ and CN are observed on and after Aug 27.24 2025, when 3I/ATLAS was within $2.7$~au of the Sun.  

To model the production rates of the strong emission features in 3I/ATLAS, we first remove contamination from reflected sunlight. Following \citet{Rahatgaonkar_2025}, \citet{Hoogendam25_KCWI}, and \citet{Hoogendam25_SNIFS}, we fit a solar analogue spectrum using the same approach and adopt the following continuum model:

\begin{equation}
    F_\mathrm{cont}(\lambda) = S \times R(\lambda) \times F_\mathrm{\odot,~analog}\left[\lambda \left(1+\frac{v}{c}\right) + \delta\lambda\right],
\end{equation}

\noindent where $S$ is the normalization factor used to convert the flux of the solar analogue to the same order of magnitude as the 3I/ATLAS spectrum, $F_\mathrm{\odot,~analog}$ is the flux of the solar analogue, $v$ is the velocity shift between the motion of 3I/ATLAS relative to earth and the solar analogue relative to earth, $\delta \lambda$ is the wavelength offset between the two spectra, and finally $R(\lambda)$ is the reflectance function which takes the form of a second order polynomial defined as

\begin{equation}
    R(\lambda) = (b_1 \lambda^2 + b_2 \lambda + 1).
\end{equation}

\noindent This model removes solar absorption features and the dust continuum, enabling us to model the emission features of 3I/ATLAS and determine the production rates of key elements and molecules.

Several emission features emerged in the spectrum of 3I/ATLAS once it had transitioned to a heliocentric distance of $r_h \leq 2.7$~au. These emission lines are associated with Ni, Fe, CN and CH. The properties of these lines are discussed in the following sections.

\subsection{Atomic Activity}

We examine prominent emission features between \(3400\) and \(4100\)~\AA, which include several atomic and molecular lines associated with outgassing Ni and Fe, as well as $\mathrm{C_3}$ \citep{Hoogendam25_KCWI, Rahatgaonkar_2025, Hutsemekers_2021}. These emission features have been detected in other ISOs \citep{Guzik_2020, Opitom_2021} and in Solar System objects \citep{Manfroid_2021, Hutsemekers_2021}. 

In the July spectra, obtained at \(r_h \approx 3.7\text{--}3.5\)~au, we do not detect atomic emission lines, indicating that 3I/ATLAS had not yet reached the level of activity required to produce strong metal emission at these distances. By 2025 Aug 27 and 2025 Nov 30, at \(r_h = 2.7\)~au and \(1.8\)~au, respectively, we detect several strong narrow emission lines from \NiI\ and \FeI\ (Figure~\ref{fig:emission_features}). While \NiI\ emission lines are clearly detected in the 2025 Aug 27.24 spectrum, the \FeI\ lines are comparatively weak, although they do seem to be present at this epoch. Such \FeI\ emission has been observed in Solar System comets at heliocentric distances out to \(3.25\)~au \citep{Manfroid_2021}. 

As noted by \citet{Rahatgaonkar_2025}, the early onset of the \NiI\ lines, at relatively large heliocentric distances with reduced solar flux, suggests that the Ni does not originate from the sublimation of metallic Ni.

If the \NiI\ emission does not arise from sublimation of pure metallic Ni, an alternative explanation is the outgassing of Ni from volatile metal carbonyl species from the surface of 3I/ATLAS \citep{Bromley21, Manfroid_2021, Rahatgaonkar_2025}. Additionally, the delay time between the onset of the \NiI\ lines relative to the \FeI\ lines can be explained if both the Ni and Fe originate from within carbonyls, such as $\mathrm{Ni(CO)_4}$ and $\mathrm{Fe(CO)_5}$ \citep{Manfroid_2021, Rahatgaonkar_2025}. The delay in \FeI\ emission would thus be the result of different sublimation temperatures of Fe carbonyl species compared to the Ni rich carbonyls. This origin for the Ni and Fe in 3I/ATLAS is supported by its evolving \NiI/\FeI\ ratio, which proved to be highly dependent on the intensity of solar flux and therefore the heliocentric distance of 3I/ATLAS \citep{Hutsemekers25}.

In addition to the Ni and Fe lines that emerge in the spectra of 3I/ATLAS just prior to perihelion, we also observe weak emission features that can be associated with $\mathrm{C_3}$ in the Nov 30$\mathrm{^{th}}$ 2025 spectrum at \(r_h \approx 1.8\)~au, as shown in Figure~\ref{fig:emission_features}. While we find weak features of $\mathrm{C_3}$, we do not identify any spectral features associated with either $\mathrm{C_2}$ or CH, which were identified in a spectrum of 3I/ATLAS measured by the Keck Cosmic Web Imager (KCWI) on Nov 16 2025 (\(r_h = 1.5\)~au) \citep{Hoogendam26_KCWI}. This is likely due to the weakness of those emission lines and the lower S/N ratio of the SNIFS spectrum relative to the higher resolution KCWI spectrum.

\subsection{CN production}

We identify several emission lines of CN in both the August and November spectra (see Figure~\ref{fig:emission_features}).
These features are noticeably absent from the earlier July spectra. We fit the CN band using a simple Haser model and adopted the parent/daughter scalelengths of $1.3\times 10^{4} / 2.1\times 10^{5} \mathrm{km}$, from \citet{AHearn:1995}. We adopted the fluorescence efficiency (g-factor; photons per molecule per second) from \citet{Schleicher:2010}. The slit width and aperture size for the August CN fit was 1.0$\prime\prime$ and 2.21$\prime\prime$ while the November slit width and aperture sizes were 0.43$\prime\prime$ and 0.84$\prime\prime$. From these fits, we find a production rate of CN on Aug 27.24, 2025, to be $\mathrm{Q(CN)} = 5.7 \pm 1.5 \times 10^{24}~\mathrm{molecules~s^{-1}}$. The CN production rate determined from the Aug 27.24 spectrum is similar to the rates reported in other work, which also observed an increase in activity as 3I/ATLAS approached the sun \citep{Rahatgaonkar_2025, Schleicher_2025ATel, SalazarManzano2025,Hoogendam25_KCWI,Hoogendam25_SNIFS}. 
Comparing the CN production rates of the other active ISO shows that
3I/ATLAS \(\left(5.7 \pm 1.5 \times 10^{24}\ \mathrm{molecules\ s^{-1}}\right)\)
exhibited a CN production rate roughly twice that of 2I/Borisov \(\left(3.7 \pm 0.4 \times 10^{24}\ \mathrm{molecules\ s^{-1}}\right)\) at a similar heliocentric distance.

In the post-perihelion observation obtained on Nov 30, 2025, the CN production rate increased to
$\mathrm{Q(CN)} = (3.2 \pm 1.2)\times 10^{25}\ \mathrm{molecules\ s^{-1}}$.
This rate is comparable to that measured with TRAPPIST-North
around the same epoch \citep{Jehin_2025}, but is nearly an order of magnitude lower than the value inferred
from Keck\,II/KCWI observations obtained \(\sim 14\) days earlier, when 3I/ATLAS was at
\(r_h = 1.5\)~au \citep{Hoogendam26_KCWI}. Thus the CN production rate appears to decrease after perihelion passage. A similar trend in the evolution of CN production was also reported from post-perihelion observations of 2I/Borisov \citep{Cordiner_2020, Prodan2024}. As 3I/ATLAS moves away from the perihelion, the production rate of CN should continue to decrease as it receives less solar flux. While the post-perihelion trend in CN production rate in both active ISOs is expected, 3I/ATLAS exhibits substantially higher CN abundance than 2I/Borisov \citep{Cordiner_2020, Prodan2024}, with $\mathrm{Q(CN)}$ differing by a factor of \(\sim 33\) at comparable heliocentric distances (\(r_h \sim 2.0\)~au).

\subsection{Dust and Coma}
At longer wavelengths, the spectrum of 3I/ATLAS is shaped by sunlight scattered by dust in the coma and by molecular emission from rovibrational bands of $\mathrm{H_2O}$, CO, and $\mathrm{CO_2}$ \citep{Cordiner_2025}. The dust scattering component is prominent in the NIR spectrum presented in this work (Figure~\ref{fig:pan_spec}), where a broad feature peaks near \(\sim 1.15\)~\micron. This feature is not apparent in the IRTF spectrum obtained on 2025 Jul 14 \citep{Yang2025}, but it is clearly present in the \textit{JWST} observation acquired $\sim 23$ days later on 2025 Aug 6 \citep{Cordiner_2025}.

Additionally, in both of the Keck-II/NIRES spectra taken on Aug 28 2025 (\(r_h = 2.6\)~au) and Nov 30 2025 (\(r_h = 1.8\)~au) we do not detect any strong absorption features associated with $\mathrm{H_2O}$-ice. This is in agreement with the weak features observed in the Jul 14 2025 IRTF spectrum \citep{Yang2025}. The lack of any deep $\mathrm{H_2O}$ absorption features as 3I/ATLAS moved through perihelion is consistent with sublimation of the previously present surface water ice, or small (micron-sized) ice grains, in agreement with other observations \citep{Cordiner_2025, Yang2025}.

\begin{figure}
    \centering
    \includegraphics[width=\linewidth, trim={0cm, 0.3cm, 0cm, 0cm}, clip=True]{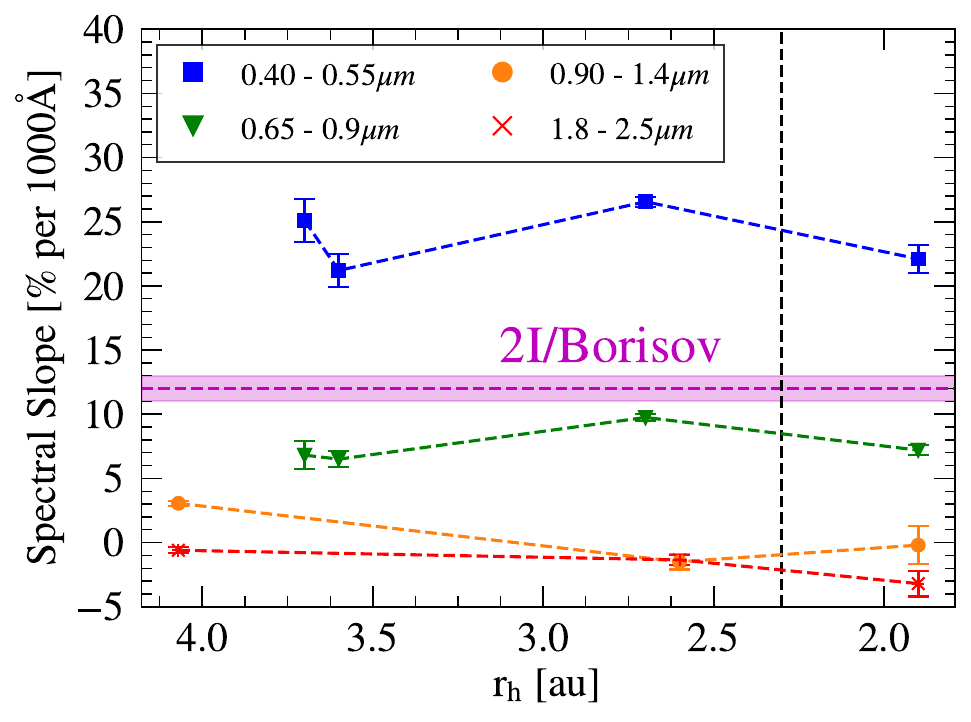}
    \caption{Evolution of the optical and NIR spectral slopes of 3I/ATLAS as it approaches perihelion. The dashed vertical line indicates heliocentric distance where 3I/ATLAS crossed perihelion. While the optical slope showed a slight increase in spectral slope between the two Keck-I/LRIS observations, the NIR showed a declining evolution between the initial NIR observation \citep{Yang2025} and the Keck-II/NIRES data. The average optical spectral slope of 2I/Borisov is also included \citep{Jewitt2023ARAA}. The spectral slope of 3I/ATLAS is similar to 2I/Borisov between $0.65 - 0.9$~\micron.}
    \label{fig:Slope_evo}
\end{figure}

\begin{figure*}
    \centering
    \includegraphics[width=\linewidth, trim={0.25cm, 0.2cm, 0.1cm, 0.2cm}, clip=True]{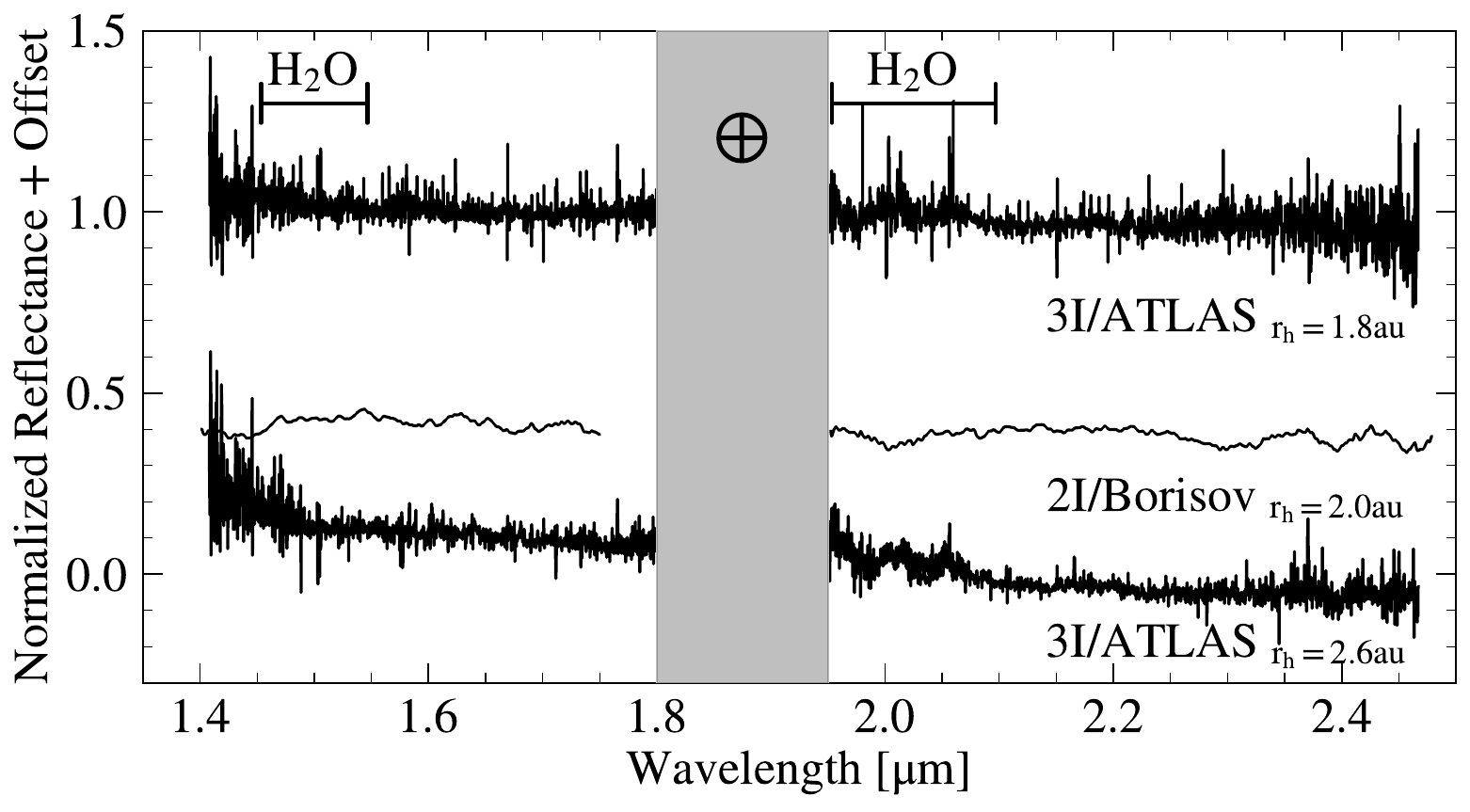}
    \caption{Comparison of 3I/ATLAS's NIR spectrum with the binned NIR spectrum of 2I/Borisov \citep{Lee_2020_2I_Perihelion} taken with FLAMINGOS-2 at a heliocentric distance of $\sim 2.0$~au. Both spectra show a negative spectral slope with no indication of the $1.5$ and $2.0$~\micron\ $\mathrm{H_2O}$ ice absorption features, which have been highlighted for clarity. The telluric region has been removed from both spectra and marked by the shaded region.}
    \label{fig:NIR_comp}
\end{figure*}

\subsection{Spectral slope}

\begin{deluxetable*}{cccc}
    \tablenum{2}
    \tablecaption{Spectral slopes determined from fitting a linear function to regions of the optical and NIR spectra}\label{tab:spec_slope}
    \tablewidth{0pt}
    \tablehead{ \colhead{UT Date} & \colhead{Spectral Region} & \colhead{$\mathrm{r_h}$} & \colhead{Spectral Slope} \\
    \colhead{} & \colhead{[\micron]} & \colhead{[au]} & \colhead{[$\%$ per 1000~\AA]} 
    }
    \startdata
    \multirow{2}{*}{26 Jul} & $0.4 - 0.55$ & 3.7 & $25.1 \pm 1.7$ \\
    & $0.65 - 0.9$ & 3.7 & $6.8 \pm 1.1$ \\
    \hline
    27 Jul & $0.4 - 0.55$ & 3.5 & $21.1 \pm 1.3$ \\
    31 Jul & $0.65 - 0.9$ & 3.5 & $6.5 \pm 0.6$ \\
    \hline
    \multirow{2}{*}{27 Aug} & $0.4 - 0.55$ & 2.7 & $26.6 \pm 0.2$ \\
    & $0.65 - 0.9$ & 2.6 & $9.6 \pm 0.3$ \\
    \multirow{2}{*}{28 Aug} & $0.9 - 1.5$ & 2.6 & $-1.5 \pm 1.1$ \\
    & $1.4 - 2.0$ & 2.6 & $-1.4 \pm 0.4$ \\
    \hline
    \multirow{4}{*}{30 Nov} & $0.4 - 0.55$ & 1.8 & $22.1 \pm 1.1$ \\
    & $0.65 - 0.9$ & 1.8 & $7.2 \pm 0.4$ \\
    & $0.9 - 1.4$ & 1.8 & $-0.2 \pm 1.5$ \\
    & $1.9 - 2.5$ & 1.8 & $-3.2 \pm 1.0$
    \enddata
\end{deluxetable*}

The shape of the reflectance spectrum of 3I/ATLAS is linked to the object's physical properties, including its surface composition and volatile driven surface activity. The shape of cometary spectra are commonly quantified using the spectral slope \citep{Licandro_2018}.
To determine the spectral slope of 3I/ATLAS, we correct for the solar continuum by dividing by a solar analogue spectrum and then fit a linear function to the resulting reflectance spectra. The uncertainty on the spectral slope is determined by combining the covariance parameters obtained from the linear fitting and spectral uncertainty in quadrature. 
The best fit spectral slopes for each wavelength region are listed in Table~\ref{tab:spec_slope}.
On 26 Jul 2025, the spectrum had a spectral slope of \(25.1 \pm 1.7\%\) per 1000~\AA\ over \(0.40\text{--}0.55\)~\micron\ and \(6.8 \pm 1.1\%\) per 1000~\AA\ over \(0.65\text{--}0.90\)~\micron. Similar spectral slopes were found in the 27/31 Jul 2025 spectra across the same regions.
By 27/28 Aug 2025, at a smaller heliocentric distance, the slopes are \(26.6 \pm 0.2\%\) per 1000~\AA\ over \(0.40\text{--}0.55\)~\micron\ and \(9.6 \pm 0.3\%\) per 1000~\AA\ over \(0.65\text{--}0.90\)~\micron, indicating no substantial change.
The spectral slopes determined from the optical observations are in general agreement with the values determined from other instruments \citep{Yang2025, Hoogendam25_KCWI, Hoogendam25_SNIFS}, with any discrepancy likely arising from the different wavelength regions used for the fitting.
Finally, the spectral slope does not show a significant change between the pre-perihelion observations and the post-perihelion observation taken on 30 Nov 2025. The post-perihelion optical data shows a decreasing spectral slope at redder wavelengths, with the region between $0.4 - 0.55$~\micron\ had a slope of \(22.1 \pm 1.1\%\) per 1000~\AA, which then decreased to \(7.2 \pm 0.4\%\) per 1000~\AA\ between $0.65 - 0.9$~\micron. The post-perihelion NIR observation on 30 Nov 2025 shows a similar trend with a spectral slope of \(-0.2 \pm 1.5\) per 1000~\AA\ between $0.9 - 1.4$~\micron\ and \(-3.2 \pm 1.0\) per 1000~\AA\ across $1.9 - 2.5$~\micron. 

The Keck\,II/NIRES Aug 28 and Nov 30, 2025 measurements show a significant change from the IRTF and VLT/X-SHOOTER observations obtained at larger heliocentric distances, \(r_h = 4.1\)~au and \(4.37\)~au, respectively \citep{Yang2025, Alvarez-Candal2025}. The IRTF NIR spectrum shows a spectral slope of \(6.7 \pm 1.4\%\), whereas the VLT/X-SHOOTER NIR spectrum is too strongly affected by telluric absorption to derive a reliable slope. The difference in NIR slope may reflect increased scattering from dust in the coma, which likely strengthened between the IRTF and NIRES epochs as 3I/ATLAS approached the Sun. Such an evolution is consistent with rising activity that enhances dust production through volatile-driven outgassing \citep{SalazarManzano2025, Cordiner_2025}. This change in the NIR spectral slope could indicate an increased contribution from pure water ice or hydrated minerals in the coma of 3I/ATLAS \citep{Yang_2009}. However, we do not detect a corresponding increase in the depths of the $\mathrm{H_2O}$ absorption features near \(1.5\) and \(2.0\)~\micron\ that were identified in the IRTF spectra, as might be expected if the later spectra were dominated by a more negative slope. 

A similar change in the spectral slope was also observed in 2I/Borisov, whose NIR spectral slope evolved from $\sim 6.0\%$ per 1000~\AA\ \citep{Yang2020} to $\sim -2.2\%$ per 1000~\AA\ \citep{Lee_2020_2I_Perihelion}. The evolution of 2I/Borisov and 3I/ATLAS occurred as they approached perihelion, although the spectral slope of 3I/ATLAS evolved much more rapidly than that of 2I/Borisov. The evolution from a positive to negative spectral slope in 3I/ATLAS occurred between a heliocentric distance of $4.1$~au \citep{Yang2025} to $2.6$~au, while 2I/Borisov transitioned as the object moved from $~2.5$~au \citep{Yang2020} to $2.0$~au \citep{Lee_2020_2I_Perihelion}. The turnover in spectral slope observed in 2I/Borisov was suggested to occur as the activity of 2I/Borisov moved from CO-dominant volatiles to $\mathrm{H_2O}$-dominant volatiles \citep{Lee_2020_2I_Perihelion}. Both 2I/Borisov and 3I/ATLAS lacked clear $\mathrm{H_2O}$ absorptions at heliocentric distances less than 3~au (Figure~\ref{fig:NIR_comp}).

\section{Conclusions}\label{sec:conclusions}

Here, we present panchromatic optical and NIR observations of the interstellar comet 3I/ATLAS. Data were obtained over the course of 6 months as 3I/ATLAS moved from a perihelion distance of 3.7 au to 1.8 au. The optical spectra of 3I/ATLAS show a drastic change between the epochs of observations, with the emergence of a large number of emission features below $0.4$~\micron. These features are associated with \NiI\ and \FeI\ lines and CN emission, along with weak features from $\mathrm{C_2}$ and $\mathrm{C_3}$. Emission from \FeI\ lines emerged in the optical spectra of 3I/ATLAS as it approached perihelion and after. From the CN emission feature, we determined a pre-perihelion CN formation rate of $5.7 \pm 1.5 \times 10^{24}~\mathrm{~molecules~s^{-1}}$ at a distance of $\Delta = 2.6$~au / $r_h = 2.7$~au. The formation rate of CN increased to $\mathrm{Q(CN)} = 3.2 \pm 1.2 \times 10^{25}~\mathrm{molecules~s^{-1}}$ after 3I/ATLAS passed perihelion, at a distance of $\Delta = 1.9$~au / $r_h = 1.8$~au. The derived CN formation rate of 3I/ATLAS was an order of magnitude larger than that of the other active interstellar object, 2I/Borisov, suggesting different formation environments, although the nuclear surface area of 3I/ATLAS may also be larger by a similar factor \citep{Hui2026}. 

We do not identify a clear absorption feature associated with $\mathrm{H_2O}$-ice at $2.0$~\micron, although the shape of the NIR spectrum around this region is consistent with a weak absorption contaminated by telluric features, making it difficult to measure a $\mathrm{H_2O}$ feature depth. We do not identify the even weaker $1.5$~\micron\ absorption feature. 
We observe a clear evolution in the NIR spectral shape from a positive slope at larger heliocentric distances to a negative slope as 3I/ATLAS moved significantly closer to the Sun. This trend was also observed in 2I/Borisov and in several solar system objects, consistent with an increase in cometary activity as 3I/ATLAS approached the sun. As 3I/ATLAS continues to move past perihelion, $\mathrm{H_2O}$ ice grains could appear in its coma again, so future NIR observations should be obtained to document its continued evolution.

\begin{acknowledgements}

We thank Brian Lemaux and the staff at Gemini North for their assistance in preparing and executing the Gemini Queue observations. We thank Alexa Anderson for helpful discussions of protoplanetary discs. 

C.A. and K.M. acknowledge support from NASA grants JWST-GO-02114, JWST-GO-02122, JWST-GO-03726, JWST-GO-04217, JWST-GO-04436, JWST-GO-04522, JWST-GO-05057, JWST-GO-05290, JWST-GO-06023, JWST-GO-06213, JWST-GO-06583, and JWST-GO-06677. Support for these programs was provided by NASA through a grant from the Space Telescope Science Institute, which is operated by the Association of Universities for Research in Astronomy, Inc., under NASA contract NAS5-03127.

W.B.H. acknowledges support from the National Science Foundation Graduate Research Fellowship Program under Grant No. 2236415. 

K.J.M., J.J.W., and A.H.\ acknowledge support from the Simons Foundation through SFI-PD-Pivot Mentor-00009672.

J.T.H. acknowledges support from NASA through the NASA Hubble Fellowship grant HST-HF2-51577.001-A, awarded by STScI. STScI is operated by the Association of Universities for Research in Astronomy, Incorporated, under NASA contract NAS5-26555.

D.O.J. acknowledges support from NSF grants AST-2407632, AST-2429450, and AST-2510993, NASA grant 80NSSC24M0023, and HST/JWST grants HST-GO-17128.028 and JWST-GO-05324.031, awarded by the Space Telescope Science Institute (STScI), which is operated by the Association of Universities for Research in Astronomy, Inc., for NASA, under contract NAS5-26555.

Based on observations obtained at the international Gemini Observatory (program GN-2025B-Q-112), a program of NSF NOIRLab, which is managed by the Association of Universities for Research in Astronomy (AURA) under a cooperative agreement with the U.S. National Science Foundation on behalf of the Gemini Observatory partnership: the U.S. National Science Foundation (United States), National Research Council (Canada), Agencia Nacional de Investigaci\'{o}n y Desarrollo (Chile), Ministerio de Ciencia, Tecnolog\'{i}a e Innovaci\'{o}n (Argentina), Minist\'{e}rio da Ci\^{e}ncia, Tecnologia, Inova\c{c}\~{o}es e Comunica\c{c}\~{o}es (Brazil), and Korea Astronomy and Space Science Institute (Republic of Korea).

Some of the data presented herein were obtained at Keck Observatory, which is a private 501(c)3 non-profit organization operated as a scientific partnership among the California Institute of Technology, the University of California, and the National Aeronautics and Space Administration. The Observatory was made possible by the generous financial support of the W. M. Keck Foundation.

This research made use of \texttt{PypeIt}\footnote{\url{https://pypeit.readthedocs.io/en/latest/}},
a Python package for semi-automated reduction of astronomical data \citep{pypeit:joss_pub, pypeit:zenodo}.
\end{acknowledgements}

\facilities{Keck-I(LRIS), Keck-II(NIRES), Gemini(GMOS), UH88(SNIFS)}

\software{astropy \citep{2013A&A...558A..33A,2018AJ....156..123A,2022ApJ...935..167A},  
          pypeit \citep{ Prochaska_2020zndo, Prochaska_2020}, spextool \citep{Vacca03, Cushing04}}




\bibliography{Bib}{}

@INCOLLECTION{Bergin2024,
       author = {{Bergin}, Edwin and {Alexander}, Conel and {Drozdovskaya}, Maria and {Gounelle}, Matthieu and {Pfalzner}, Susanne},
        title = "{Interstellar Heritage and the Birth Environment of the Solar System}",
    booktitle = {Comets III},
         year = 2024,
       editor = {{Meech}, Karen. J. and {Combi}, Michael. R. and {Bockel{\'e}e-Morvan}, Dominique and {Raymond}, Sean. N. and {Zolensky}, Michael. E.},
        pages = {3-32},
          doi = {10.2458/azu_uapress_9780816553631-ch001},
       adsurl = {https://ui.adsabs.harvard.edu/abs/2024come.book....3B}
}

@ARTICLE{Seligman2024,
       author = {{Seligman}, Darryl Z. and {Farnocchia}, Davide and {Micheli}, Marco and {Hainaut}, Olivier R. and {Hsieh}, Henry H. and {Feinstein}, Adina D. and {Chesley}, Steven R. and {Taylor}, Aster G. and {Masiero}, Joseph and {Meech}, Karen J.},
        title = "{Two distinct populations of dark comets delineated by orbits and sizes}",
      journal = {Proceedings of the National Academy of Science},
         year = 2024,
        month = dec,
       volume = {121},
       number = {51},
          eid = {e2406424121},
        pages = {e2406424121},
          doi = {10.1073/pnas.2406424121},
archivePrefix = {arXiv},
       eprint = {2412.07603},
 primaryClass = {astro-ph.EP},
       adsurl = {https://ui.adsabs.harvard.edu/abs/2024PNAS..12106424S}
}

@INCOLLECTION{Meech2020,
       author = {{Meech}, K. and {Raymond}, S.~N.},
        title = "{Origin of Earth's Water: Sources and Constraints}",
    booktitle = {Planetary Astrobiology},
         year = 2020,
       editor = {{Meadows}, Victoria S. and {Arney}, Giada N. and {Schmidt}, Britney E. and {Des Marais}, David J.},
        pages = {325},
          doi = {10.2458/azu_uapress_9780816540068},
       adsurl = {https://ui.adsabs.harvard.edu/abs/2020plas.book..325M}
}

@ARTICLE{Belyakov26,
       author = {{Belyakov}, Matthew and {Wong}, Ian and {Bolin}, Bryce T. and {Ryleigh Davis}, M. and {Bromley}, Steven J. and {Lisse}, Carey M. and {Brown}, Michael E.},
        title = "{The Volatile Inventory of 3I/ATLAS as seen with JWST/MIRI}",
      journal = {arXiv e-prints},
         year = 2026,
        month = jan,
          eid = {arXiv:2601.22034},
        pages = {arXiv:2601.22034},
          doi = {10.48550/arXiv.2601.22034},
archivePrefix = {arXiv},
       eprint = {2601.22034},
 primaryClass = {astro-ph.EP},
       adsurl = {https://ui.adsabs.harvard.edu/abs/2026arXiv260122034B}
}

@ARTICLE{Tan26,
       author = {{Tan}, Hanjie and {Yan}, Xiaoran and {Li}, Jian-Yang},
        title = "{Perihelion Asymmetry in the Water Production Rate of the Interstellar Object 3I/ATLAS}",
      journal = {arXiv e-prints},
         year = 2026,
        month = jan,
          eid = {arXiv:2601.15443},
        pages = {arXiv:2601.15443},
          doi = {10.48550/arXiv.2601.15443},
archivePrefix = {arXiv},
       eprint = {2601.15443},
 primaryClass = {astro-ph.EP},
       adsurl = {https://ui.adsabs.harvard.edu/abs/2026arXiv260115443T}
}

@ARTICLE{Ciesla14,
       author = {{Ciesla}, Fred J.},
        title = "{The Phases of Water Ice in the Solar Nebula}",
      journal = {\apjl},
         year = 2014,
        month = mar,
       volume = {784},
       number = {1},
          eid = {L1},
        pages = {L1},
          doi = {10.1088/2041-8205/784/1/L1},
archivePrefix = {arXiv},
       eprint = {1402.5333},
 primaryClass = {astro-ph.EP},
       adsurl = {https://ui.adsabs.harvard.edu/abs/2014ApJ...784L...1C}
}

@ARTICLE{Muller21,
       author = {{M{\"u}ller}, Jonas and {Savvidou}, Sofia and {Bitsch}, Bertram},
        title = "{The water-ice line as a birthplace of planets: implications of a species-dependent dust fragmentation threshold}",
      journal = {\aap},
         year = 2021,
        month = jun,
       volume = {650},
          eid = {A185},
        pages = {A185},
          doi = {10.1051/0004-6361/202039930},
archivePrefix = {arXiv},
       eprint = {2104.06749},
 primaryClass = {astro-ph.EP},
       adsurl = {https://ui.adsabs.harvard.edu/abs/2021A&A...650A.185M}
}

@ARTICLE{Bitsch16,
       author = {{Bitsch}, Bertram and {Johansen}, Anders},
        title = "{Influence of the water content in protoplanetary discs on planet migration and formation}",
      journal = {\aap},
         year = 2016,
        month = may,
       volume = {590},
          eid = {A101},
        pages = {A101},
          doi = {10.1051/0004-6361/201527676},
archivePrefix = {arXiv},
       eprint = {1603.01125},
 primaryClass = {astro-ph.EP},
       adsurl = {https://ui.adsabs.harvard.edu/abs/2016A&A...590A.101B}
}

@ARTICLE{Chyba_1990a,
       author = {{Chyba}, C.~F.},
        title = "{Impact delivery and erosion of planetary oceans in the early inner Solar System}",
      journal = {\nat},
         year = 1990,
        month = jan,
       volume = {343},
       number = {6254},
        pages = {129-133},
          doi = {10.1038/343129a0},
       adsurl = {https://ui.adsabs.harvard.edu/abs/1990Natur.343..129C}
}

@ARTICLE{Mottl07,
       author = {{Mottl}, M. and {Glazer}, B. and {Kaiser}, R. and {Meech}, K.},
        title = "{Water and astrobiology}",
      journal = {Chemie der Erde / Geochemistry},
         year = 2007,
        month = dec,
       volume = {67},
       number = {4},
        pages = {253-282},
          doi = {10.1016/j.chemer.2007.09.002},
       adsurl = {https://ui.adsabs.harvard.edu/abs/2007ChEG...67..253M}
}

@ARTICLE{Hoogendam26_KCWI,
       author = {{Hoogendam}, Willem B. and {Jones}, David O. and {Yang}, Bin and {Shappee}, Benjamin J. and {Wray}, James J. and {Meech}, Karen J. and {Ashall}, Christopher and {Desai}, Dhvanil D. and {Hinkle}, Jason T. and {Hoffman}, Andrew M. and {Medler}, Kyle and {Pfeffer}, Cameron and {Zhao}, Ruining},
        title = "{Post-Perihelion Integral Field Spectroscopy of the Interstellar Comet 3I/ATLAS}",
      journal = {arXiv e-prints},
         year = 2026,
        month = jan,
          eid = {arXiv:2601.16983},
        pages = {arXiv:2601.16983},
archivePrefix = {arXiv},
       eprint = {2601.16983},
 primaryClass = {astro-ph.EP},
       adsurl = {https://ui.adsabs.harvard.edu/abs/2026arXiv260116983H}
}

@ARTICLE{Vacca03,
       author = {{Vacca}, William D. and {Cushing}, Michael C. and {Rayner}, John T.},
        title = "{A Method of Correcting Near-Infrared Spectra for Telluric Absorption}",
      journal = {\pasp},
         year = 2003,
        month = mar,
       volume = {115},
       number = {805},
        pages = {389-409},
          doi = {10.1086/346193},
archivePrefix = {arXiv},
       eprint = {astro-ph/0211255},
 primaryClass = {astro-ph},
       adsurl = {https://ui.adsabs.harvard.edu/abs/2003PASP..115..389V}
}

@ARTICLE{Cushing04,
       author = {{Cushing}, Michael C. and {Vacca}, William D. and {Rayner}, John T.},
        title = "{Spextool: A Spectral Extraction Package for SpeX, a 0.8-5.5 Micron Cross-Dispersed Spectrograph}",
      journal = {\pasp},
         year = 2004,
        month = apr,
       volume = {116},
       number = {818},
        pages = {362-376},
          doi = {10.1086/382907},
       adsurl = {https://ui.adsabs.harvard.edu/abs/2004PASP..116..362C}
}

@ARTICLE{Medler25_HISS,
       author = {{Medler}, K. and {Ashall}, C. and {Shahbandeh}, M. and {DerKacy}, J.~M. and {Hoogendam}, W.~B. and {Jones}, D.~O. and {Shappee}, B.~J. and {Hinkle}, J.~T. and {Pfeffer}, C.~M. and {Baron}, E. and {Hoeflich}, P. and {Hsiao}, E.},
        title = "{The Hawaii Infrared Supernova Study (HISS): Spectroscopic Data Release 1}",
      journal = {arXiv e-prints},
         year = 2025,
        month = may,
          eid = {arXiv:2505.18507},
        pages = {arXiv:2505.18507},
          doi = {10.48550/arXiv.2505.18507},
archivePrefix = {arXiv},
       eprint = {2505.18507},
 primaryClass = {astro-ph.HE},
       adsurl = {https://ui.adsabs.harvard.edu/abs/2025arXiv250518507M}
}

@ARTICLE{Hook_2004,
       author = {{Hook}, I.~M. and {J{\o}rgensen}, Inger and {Allington-Smith}, J.~R. and {Davies}, R.~L. and {Metcalfe}, N. and {Murowinski}, R.~G. and {Crampton}, D.},
        title = "{The Gemini-North Multi-Object Spectrograph: Performance in Imaging, Long-Slit, and Multi-Object Spectroscopic Modes}",
      journal = {\pasp},
         year = 2004,
        month = may,
       volume = {116},
       number = {819},
        pages = {425-440},
          doi = {10.1086/383624},
       adsurl = {https://ui.adsabs.harvard.edu/abs/2004PASP..116..425H}
}

@ARTICLE{SalazarManzano2025,
       author = {{Salazar Manzano}, Luis E. and {Lin}, Hsing Wen and {Taylor}, Aster G. and {Seligman}, Darryl Z. and {Adams}, Fred C. and {Gerdes}, David W. and {Ruch}, Thomas and {Frincke}, Tessa T. and {Napier}, Kevin J.},
        title = "{Onset of CN Emission in 3I/ATLAS: Evidence for Strong Carbon-Chain Depletion}",
      journal = {arXiv e-prints},
         year = 2025,
        month = sep,
          eid = {arXiv:2509.01647},
        pages = {arXiv:2509.01647},
          doi = {10.48550/arXiv.2509.01647},
archivePrefix = {arXiv},
       eprint = {2509.01647},
 primaryClass = {astro-ph.EP},
       adsurl = {https://ui.adsabs.harvard.edu/abs/2025arXiv250901647S}
}

@ARTICLE{Hoogendam25_KCWI,
       author = {{Hoogendam}, W.~B. and {Shappee}, B.~J. and {Wray}, J.~J. and {Yang}, B. and {Meech}, K.~J. and {Ashall}, C. and {Desai}, D.~D. and {Hart}, K. and {Hinkle}, J.~T. and {Hoffman}, A. and {Hu}, E.~M. and {Jones}, D.~O. and {Medler}, K.},
        title = "{Spatial Profiles of 3I/ATLAS CN and Ni Outgassing from Keck/KCWI Integral Field Spectroscopy}",
      journal = {arXiv e-prints},
         year = 2025,
        month = oct,
          eid = {arXiv:2510.11779},
        pages = {arXiv:2510.11779},
archivePrefix = {arXiv},
       eprint = {2510.11779},
 primaryClass = {astro-ph.EP},
       adsurl = {https://ui.adsabs.harvard.edu/abs/2025arXiv251011779H}
}

@ARTICLE{Hoogendam25_SNIFS,
       author = {{Hoogendam}, W.~B. and {Kuesters}, D. and {Shappee}, B.~J. and {Aldering}, G. and {Wray}, J.~J. and {Yang}, B. and {Meech}, K.~J. and {Tucker}, M.~A. and {Huber}, M.~E. and {Auchettl}, K. and {Angus}, C.~R. and {Desai}, D.~D. and {Hinkle}, J.~T. and {Kiyokawa}, J. and {Paek}, G.~S.~H. and {Romagnoli}, S. and {Shi}, J. and {Syncatto}, A. and {Ashall}, C. and {Dixon}, M. and {Hart}, K. and {Hoffman}, A.~M. and {Jones}, D.~O. and {Medler}, K. and {Pfeffer}, C.},
        title = "{University of Hawaii 88-inch Telescope Observations of the Interstellar Comet 3I/ATLAS: Spectrophotometric Blue-Sensitive Spectral Time Series Spanning Two Months from Discovery}",
      journal = {arXiv e-prints},
         year = 2025,
        month = dec,
          eid = {arXiv:2512.09020},
        pages = {arXiv:2512.09020},
          doi = {10.48550/arXiv.2512.09020},
archivePrefix = {arXiv},
       eprint = {2512.09020},
 primaryClass = {astro-ph.EP},
       adsurl = {https://ui.adsabs.harvard.edu/abs/2025arXiv251209020H}
}

@ARTICLE{Hinkle25_JCMT,
       author = {{Hinkle}, Jason T. and {Yang}, Bin and {Meech}, Karen J. and {Hoffman}, Andrew and {Shappee}, Benjamin J. and {Hoogendam}, W.~B. and {Wray}, James J.},
        title = "{JCMT Constraints on the Early-Time HCN and CO Emission and HCN Temporal Evolution of 3I/ATLAS}",
      journal = {arXiv e-prints},
         year = 2025,
        month = dec,
          eid = {arXiv:2512.02106},
        pages = {arXiv:2512.02106},
          doi = {10.48550/arXiv.2512.02106},
archivePrefix = {arXiv},
       eprint = {2512.02106},
 primaryClass = {astro-ph.EP},
       adsurl = {https://ui.adsabs.harvard.edu/abs/2025arXiv251202106H}
}

@ARTICLE{Lisse_2025,
       author = {{Lisse}, Carey M. and {Bach}, Yoonsoo P. and {Bryan}, Sean A. and {Crill}, Brendan P. and {Korngut}, Phil M. and {Cukierman}, Ari J. and {Werner}, Michael W. and {Cooray}, Asantha and {Zemcov}, Michael and {Tolls}, Volker and {Melnick}, Gary J. and {Faisst}, Andreas L. and {Dowell}, C. Darren and {Choi}, Seungwon and {Geem}, Jooyeon and {Ishiguro}, Masateru and {Jo}, Hangbin and {Lim}, Bumhoo and {Mahlke}, Max and {Hora}, Joseph L. and {Cheng}, Yun-Ting and {Everett}, Spencer and {Lee}, Jeong-Eun and {Rustamkulov}, Zafar and {Jin}, Sunho and {Hui}, Howard and {Masters}, Daniel C. and {Nguyen}, Chi H. and {Paladini}, Roberta and {Yang}, Yujin and {Bock}, James J. and {Dor{\'e}}, O. and {Sitko}, M.~L. and {Champagne}, C. and {Connelley}, M. and {Emery}, J.~P. and {Fernandez}, Y.~R. and {Reach}, W.~T.},
        title = "{SPHEREx Pre-Perihelion Mapping of $\mathrm{H_2O}$, $\mathrm{CO_2}$, and $\mathrm{CO}$ in Interstellar Object 3I/ATLAS}",
      journal = {arXiv e-prints},
         year = 2025,
        month = dec,
          eid = {arXiv:2512.07318},
        pages = {arXiv:2512.07318},
archivePrefix = {arXiv},
       eprint = {2512.07318},
 primaryClass = {astro-ph.EP},
       adsurl = {https://ui.adsabs.harvard.edu/abs/2025arXiv251207318L}
}

@ARTICLE{Labrie_DRAGONS_2023,
       author = {{Labrie}, K. and {Simpson}, C. and {Cardenes}, R. and {Turner}, J. and {Soraisam}, M. and {Quint}, B. and {Oberdorf}, O. and {Placco}, V.~M. and {Berke}, D. and {Smirnova}, O. and {Conseil}, S. and {Vacca}, W.~D. and {Thomas-Osip}, J.},
        title = "{DRAGONS-A Quick Overview}",
      journal = {Research Notes of the American Astronomical Society},
         year = 2023,
        month = oct,
       volume = {7},
       number = {10},
          eid = {214},
        pages = {214},
          doi = {10.3847/2515-5172/ad0044},
archivePrefix = {arXiv},
       eprint = {2310.03048},
 primaryClass = {astro-ph.IM},
       adsurl = {https://ui.adsabs.harvard.edu/abs/2023RNAAS...7..214L}
}

@software{Labrie_DRAGONS_2023zndo,
       author = {{Labrie}, Kathleen and {Simpson}, Chris and {Turner}, James and {Quint}, Bruno and {Conseil}, Simon and {Oberdorf}, Oliver and {Soraisam}, Monika and {Placco}, Vinicius and {Smirnova}, Olesja and {Berke}, Daniel and {Vacca}, William},
        title = "{DRAGONS}",
         year = 2023,
        month = apr,
          eid = {10.5281/zenodo.7776065},
          doi = {10.5281/zenodo.7776065},
      version = {3.1.0},
    publisher = {Zenodo},
       adsurl = {https://ui.adsabs.harvard.edu/abs/2023zndo...7776065L}
}

@ARTICLE{Hutsemekers_2021,
       author = {{Hutsem{\'e}kers}, D. and {Manfroid}, J. and {Jehin}, E. and {Opitom}, C. and {Moulane}, Y.},
        title = "{FeI and NiI in cometary atmospheres. Connections between the NiI/FeI abundance ratio and chemical characteristics of Jupiter-family and Oort-cloud comets}",
      journal = {\aap},
         year = 2021,
        month = aug,
       volume = {652},
          eid = {L1},
        pages = {L1},
          doi = {10.1051/0004-6361/202141554},
archivePrefix = {arXiv},
       eprint = {2107.05932},
 primaryClass = {astro-ph.EP},
       adsurl = {https://ui.adsabs.harvard.edu/abs/2021A&A...652L...1H}
}

@ARTICLE{Schleicher_2025ATel,
       author = {{Schleicher}, David},
        title = "{The Detection of CN in Interstellar Comet 3I/ATLAS}",
      journal = {The Astronomer's Telegram},
         year = 2025,
        month = aug,
       volume = {17352},
        pages = {1},
       adsurl = {https://ui.adsabs.harvard.edu/abs/2025ATel17352....1S}
}

@ARTICLE{Coulson25,
       author = {{Coulson}, Iain M. and {Kuan}, Yi-Jehng and {Charnley}, Steven B. and {Cordiner}, Martin A. and {Chuang}, Yo-Ling and {Lee}, Yueh-Ning and {Lin}, Min-Kai and {Milam}, Stefanie N. and {Pimpanuwat}, Bannawit and {Roth}, Nathan X. and {{\.Z}{\'o}{\l}towski}, Micha{\l}},
        title = "{JCMT detection of HCN emission from 3I/ATLAS at 2.1 AU}",
      journal = {arXiv e-prints},
         year = 2025,
        month = oct,
          eid = {arXiv:2510.02817},
        pages = {arXiv:2510.02817},
          doi = {10.48550/arXiv.2510.02817},
archivePrefix = {arXiv},
       eprint = {2510.02817},
 primaryClass = {astro-ph.EP},
       adsurl = {https://ui.adsabs.harvard.edu/abs/2025arXiv251002817C}
}

@ARTICLE{Jehin_2025,
       author = {{Jehin}, E. and {Hmiddouch}, S. and {Aravind}, K. and {Manfroid}, J. and {Benkhaldoun}, Z. and {Jabiri}, A.},
        title = "{TRAPPIST first post-perihelion production rates of the Interstellar comet 3I/ATLAS}",
      journal = {The Astronomer's Telegram},
         year = 2025,
        month = dec,
       volume = {17515},
        pages = {1},
       adsurl = {https://ui.adsabs.harvard.edu/abs/2025ATel17515....1J}
}

@ARTICLE{Prodan2024,
       author = {{Prodan}, George P. and {Popescu}, Marcel and {Licandro}, Javier and {Akhlaghi}, Mohammad and {de Le{\'o}n}, Julia and {Tatsumi}, Eri and {Pastrav}, Bogdan Adrian and {Hibbert}, Jacob M. and {V{\v{a}}duvescu}, Ovidiu and {Simion}, Nicolae Gabriel and {Pall{\'e}}, Enric and {Narita}, Norio and {Fukui}, Akihiko and {Murgas}, Felipe},
        title = "{Pre-perihelion monitoring of interstellar comet 2I/Borisov}",
      journal = {\mnras},
         year = 2024,
        month = apr,
       volume = {529},
       number = {4},
        pages = {3521-3535},
          doi = {10.1093/mnras/stae539},
archivePrefix = {arXiv},
       eprint = {2402.12428},
 primaryClass = {astro-ph.EP},
       adsurl = {https://ui.adsabs.harvard.edu/abs/2024MNRAS.529.3521P}
}

@ARTICLE{Yang2020,
       author = {{Yang}, Bin and {Kelley}, Michael S.~P. and {Meech}, Karen J. and {Keane}, Jacqueline V. and {Protopapa}, Silvia and {Bus}, Schelte J.},
        title = "{Searching for water ice in the coma of interstellar object 2I/Borisov}",
      journal = {\aap},
         year = 2020,
        month = feb,
       volume = {634},
          eid = {L6},
        pages = {L6},
          doi = {10.1051/0004-6361/201937129},
archivePrefix = {arXiv},
       eprint = {1912.05318},
 primaryClass = {astro-ph.EP},
       adsurl = {https://ui.adsabs.harvard.edu/abs/2020A&A...634L...6Y}
}

@ARTICLE{AHearn:1995,
       author = {{A'Hearn}, Michael F. and {Millis}, Robert C. and {Schleicher}, David O. and {Osip}, David J. and {Birch}, Peter V.},
        title = "{The ensemble properties of comets: Results from narrowband photometry of 85 comets, 1976-1992.}",
      journal = {\icarus},
         year = 1995,
        month = dec,
       volume = {118},
       number = {2},
        pages = {223-270},
          doi = {10.1006/icar.1995.1190},
       adsurl = {https://ui.adsabs.harvard.edu/abs/1995Icar..118..223A}
}

@ARTICLE{Oke:1995,
       author = {{Oke}, J.~B. and {Cohen}, J.~G. and {Carr}, M. and {Cromer}, J. and {Dingizian}, A. and {Harris}, F.~H. and {Labrecque}, S. and {Lucinio}, R. and {Schaal}, W. and {Epps}, H. and {Miller}, J.},
        title = "{The Keck Low-Resolution Imaging Spectrometer}",
      journal = {\pasp},
         year = 1995,
        month = apr,
       volume = {107},
        pages = {375},
          doi = {10.1086/133562},
       adsurl = {https://ui.adsabs.harvard.edu/abs/1995PASP..107..375O}
}

@INPROCEEDINGS{McCarthy:1998,
       author = {{McCarthy}, James K. and {Cohen}, Judith G. and {Butcher}, Brad and {Cromer}, John and {Croner}, Ernest and {Douglas}, William R. and {Goeden}, Richard M. and {Grewal}, Tony and {Lu}, Barry and {Petrie}, Harold L. and {Weng}, Tianxiang and {Weber}, Bob and {Koch}, Donald G. and {Rodgers}, John M.},
        title = "{Blue channel of the Keck low-resolution imaging spectrometer}",
    booktitle = {Optical Astronomical Instrumentation},
         year = 1998,
       editor = {{D'Odorico}, Sandro},
       series = {Society of Photo-Optical Instrumentation Engineers (SPIE) Conference Series},
       volume = {3355},
        month = jul,
        pages = {81-92},
          doi = {10.1117/12.316831},
       adsurl = {https://ui.adsabs.harvard.edu/abs/1998SPIE.3355...81M}
}

@software{Prochaska_2020zndo,
       author = {{Prochaska}, J. Xavier and {Hennawi}, Joseph and {Cooke}, Ryan and {Westfall}, Kyle and {Wang}, Feige and {EmAstro} and {Tiffanyhsyu} and {Wasserman}, Asher and {Villaume}, Alexa and {Marijana777} and {Schindler}, JT and {Young}, David and {Simha}, Sunil and {Wilde}, Matt and {Tejos}, Nicolas and {Isbell}, Jacob and {Fl{\"o}rs}, Andreas and {Sandford}, Nathan and {Vasovi{\'c}}, Zlatan and {Betts}, Edward and {Holden}, Brad},
        title = "{pypeit/PypeIt: Release 1.0.0}",
         year = 2020,
        month = apr,
          eid = {10.5281/zenodo.3743493},
          doi = {10.5281/zenodo.3743493},
      version = {v1.0.0},
    publisher = {Zenodo},
       adsurl = {https://ui.adsabs.harvard.edu/abs/2020zndo...3743493P}
}

@ARTICLE{Prochaska_2020,
       author = {{Prochaska}, J. and {Hennawi}, Joseph and {Westfall}, Kyle and {Cooke}, Ryan and {Wang}, Feige and {Hsyu}, Tiffany and {Davies}, Frederick and {Farina}, Emanuele and {Pelliccia}, Debora},
        title = "{PypeIt: The Python Spectroscopic Data Reduction Pipeline}",
      journal = {The Journal of Open Source Software},
         year = 2020,
        month = dec,
       volume = {5},
       number = {56},
          eid = {2308},
        pages = {2308},
          doi = {10.21105/joss.02308},
archivePrefix = {arXiv},
       eprint = {2005.06505},
 primaryClass = {astro-ph.IM},
       adsurl = {https://ui.adsabs.harvard.edu/abs/2020JOSS....5.2308P}
}

@ARTICLE{Hutsemekers25,
       author = {{Hutsem{\'e}kers}, Damien and {Manfroid}, Jean and {Jehin}, Emmanu{\"e}l and {Opitom}, Cyrielle and {Bannister}, Michele and {Carvajal}, Juan Pablo and {Dorsey}, Rosemary and {Aravind}, K and {Luco}, Baltasar and {Murphy}, Brian and {Puzia}, Thomas H. and {Rahatgaonkar}, Rohan},
        title = "{Extreme NiI/FeI abundance ratio in the coma of the interstellar comet 3I/ATLAS}",
      journal = {arXiv e-prints},
         year = 2025,
        month = sep,
          eid = {arXiv:2509.26053},
        pages = {arXiv:2509.26053},
          doi = {10.48550/arXiv.2509.26053},
archivePrefix = {arXiv},
       eprint = {2509.26053},
 primaryClass = {astro-ph.EP},
       adsurl = {https://ui.adsabs.harvard.edu/abs/2025arXiv250926053H}
}

@ARTICLE{Hoogendam25_epr,
       author = {{Hoogendam}, W.~B. and {Jones}, D.~O. and {Ashall}, C. and {Shappee}, B.~J. and {Foley}, R.~J. and {Tucker}, M.~A. and {Huber}, M.~E. and {Auchettl}, K. and {Desai}, D.~D. and {Do}, A. and {Hinkle}, J.~T. and {Romagnoli}, S. and {Shi}, J. and {Syncatto}, A. and {Angus}, C.~R. and {Chambers}, K.~C. and {Coulter}, D.~A. and {Davis}, K.~W. and {de Boer}, T. and {Gagliano}, A. and {Kong}, M. and {Lin}, C. -C. and {Lowe}, T.~B. and {Magnier}, E.~A. and {Minguez}, P. and {Pan}, Y. -C. and {Patra}, K.~C. and {Severson}, S.~A. and {Taggart}, K. and {Wasserman}, A.~R. and {Yadavalli}, S.~K. and {Chen}, P. and {Post}, R.~S.},
        title = "{Seeing the Outer Edge of the Infant Type Ia Supernova 2024epr in the Optical and Near Infrared}",
      journal = {The Open Journal of Astrophysics},
         year = 2025,
        month = aug,
       volume = {8},
          eid = {120},
        pages = {120},
          doi = {10.33232/001c.143462},
archivePrefix = {arXiv},
       eprint = {2502.17556},
 primaryClass = {astro-ph.HE},
       adsurl = {https://ui.adsabs.harvard.edu/abs/2025OJAp....8E.120H}
}

@ARTICLE{Hoogendam25_pxl,
       author = {{Hoogendam}, W.~B. and {Ashall}, C. and {Jones}, D.~O. and {Shappee}, B.~J. and {Tucker}, M.~A. and {Huber}, M.~E. and {Auchettl}, K. and {Desai}, D.~D. and {Do}, A. and {Hinkle}, J.~T. and {Kong}, M.~Y. and {Romagnoli}, S. and {Shi}, J. and {Syncatto}, A. and {Kilpatrick}, C.~D.},
        title = "{Early and Extensive Ultraviolet through Near Infrared Observations of the Intermediate-luminosity Type Iax Supernovae 2024pxl}",
      journal = {\apj},
         year = 2025,
        month = aug,
       volume = {988},
       number = {2},
          eid = {209},
        pages = {209},
          doi = {10.3847/1538-4357/ade787},
archivePrefix = {arXiv},
       eprint = {2505.04610},
 primaryClass = {astro-ph.HE},
       adsurl = {https://ui.adsabs.harvard.edu/abs/2025ApJ...988..209H}
}

@INPROCEEDINGS{Aldering_2007,
       author = {{Aldering}, Gregory S. and {Supernova Factory}, Nearby},
        title = "{The SuperNova Integral Field Spectrograph}",
    booktitle = {American Astronomical Society Meeting Abstracts \#210},
         year = 2007,
       series = {American Astronomical Society Meeting Abstracts},
       volume = {210},
        month = may,
          eid = {82.07},
        pages = {82.07},
       adsurl = {https://ui.adsabs.harvard.edu/abs/2007AAS...210.8207A}
}

@ARTICLE{Bromley21,
       author = {{Bromley}, S.~J. and {Neff}, B. and {Loch}, S.~D. and {Marler}, J.~P. and {Orsz{\'a}gh}, J. and {Venkataramani}, K. and {Bodewits}, D.},
        title = "{Atomic Iron and Nickel in the Coma of C/1996 B2 (Hyakutake): Production Rates, Emission Mechanisms, and Possible Parents}",
      journal = {\psj},
         year = 2021,
        month = dec,
       volume = {2},
       number = {6},
          eid = {228},
        pages = {228},
          doi = {10.3847/PSJ/ac2dff},
archivePrefix = {arXiv},
       eprint = {2106.04701},
 primaryClass = {astro-ph.EP},
       adsurl = {https://ui.adsabs.harvard.edu/abs/2021PSJ.....2..228B}
}

@ARTICLE{Tonry2025,
       author = {{Tonry}, John and {Denneau}, Larry and {Alarcon}, Miguel and {Clocchiatti}, Alejandro and {Erasmus}, Nicolas and {Fitzsimmons}, Alan and {Licandro}, Javier and {Meech}, Karen and {Siverd}, Robert and {Weiland}, Henry},
        title = "{ATLAS Photometry of Interstellar Object 3I/ATLAS}",
      journal = {arXiv e-prints},
         year = 2025,
        month = sep,
          eid = {arXiv:2509.05562},
        pages = {arXiv:2509.05562},
          doi = {10.48550/arXiv.2509.05562},
archivePrefix = {arXiv},
       eprint = {2509.05562},
 primaryClass = {astro-ph.EP},
       adsurl = {https://ui.adsabs.harvard.edu/abs/2025arXiv250905562T}
}

@ARTICLE{Xing2025,
       author = {{Xing}, Zexi and {Oset}, Shawn and {Noonan}, John and {Bodewits}, Dennis},
        title = "{Water Detection in the Interstellar Object 3I/ATLAS}",
      journal = {arXiv e-prints},
         year = 2025,
        month = aug,
          eid = {arXiv:2508.04675},
        pages = {arXiv:2508.04675},
          doi = {10.48550/arXiv.2508.04675},
archivePrefix = {arXiv},
       eprint = {2508.04675},
 primaryClass = {astro-ph.EP},
       adsurl = {https://ui.adsabs.harvard.edu/abs/2025arXiv250804675X}
}

@ARTICLE{Lisse2025,
       author = {{Lisse}, C.~M. and {Bach}, Y.~P. and {Bryan}, S. and {Crill}, B.~P. and {Cukierman}, A. and {Dor{\'e}}, O. and {Fabinsky}, B. and {Faisst}, A. and {Korngut}, P.~M. and {Melnick}, G. and {Rustamkulov}, Z. and {Tolls}, V. and {Werner}, M. and {Sitko}, M.~L. and {Champagne}, C. and {Connelley}, M. and {Emery}, J.~P. and {Fernandez}, Y.~R. and {Yang}, B. and {the SPHEREx Science Team}},
        title = "{SPHEREx Discovery of Strong Water Ice Absorption and an Extended Carbon Dioxide Coma in 3I/ATLAS}",
      journal = {arXiv e-prints},
         year = 2025,
        month = aug,
          eid = {arXiv:2508.15469},
        pages = {arXiv:2508.15469},
          doi = {10.48550/arXiv.2508.15469},
archivePrefix = {arXiv},
       eprint = {2508.15469},
 primaryClass = {astro-ph.EP},
       adsurl = {https://ui.adsabs.harvard.edu/abs/2025arXiv250815469L}
}

@ARTICLE{Chandler2025,
       author = {{Chandler}, Colin Orion and {Bernardinelli}, Pedro H. and {Juri{\'c}}, Mario and {Singh}, Devanshi and {Hsieh}, Henry H. and {Sullivan}, Ian and {Jones}, R. Lynne and {Kurlander}, Jacob A. and {Vavilov}, Dmitrii and {Eggl}, Siegfried and {Holman}, Matthew and {Spoto}, Federica and {Schwamb}, Megan E. and {Christensen}, Eric J. and {Beebe}, Wilson and {Roodman}, Aaron and {Lim}, Kian-Tat and {Jenness}, Tim and {Bosch}, James and {Smart}, Brianna and {Bellm}, Eric and {MacBride}, Sean and {Rawls}, Meredith L. and {Greenstreet}, Sarah and {Slater}, Colin and {Heinze}, Aren and {Ivezi{\'c}}, {\v{Z}}eljko and {Blum}, Bob and {Connolly}, Andrew and {Daues}, Gregory and {Makadia}, Rahil and {Gower}, Michelle and {Bryce Kalmbach}, J. and {Monet}, David and {Bannister}, Michele T. and {Dones}, Luke and {Dorsey}, Rosemary C. and {Fraser}, Wesley C. and {Forbes}, John C. and {Fuentes}, Cesar and {Holt}, Carrie E. and {Inno}, Laura and {Jones}, Geraint H. and {Knight}, Matthew M. and {Lintott}, Chris J. and {Lister}, Tim and {Lupton}, Robert and {Mendoza Magbanua}, Mark Jesus and {Malhotra}, Renu and {Mueller}, Beatrice E.~A. and {Murtagh}, Joseph and {Pandey}, Nitya and {Reach}, William T. and {Samarasinha}, Nalin H. and {Seligman}, Darryl Z. and {Snodgrass}, Colin and {Solontoi}, Michael and {Szab{\'o}}, Gyula M. and {White}, Ellie and {Womack}, Maria and {Young}, Leslie A. and {Allbery}, Russ and {Armellin}, Roberto and {Aubourg}, {\'E}ric and {Avdellidou}, Chrysa and {Azfar}, Farrukh and {Bauer}, James and {Bechtol}, Keith and {Belyakov}, Matthew and {Benecchi}, Susan D. and {Bertini}, Ivano and {Bolin}, Bryce T. and {Bose}, vMaitrayee and {Buchanan}, Laura E. and {Boucaud}, Alexandre and {Boufleur}, Rodrigo C. and {Boutigny}, Dominique and {Braga-Ribas}, Felipe and {Calabrese}, Daniel and {Camargo}, J.~I.~B. and {Caplar}, Neven and {Carry}, Benoit and {Carvajal}, Juan Pablo and {Choi}, Yumi and {Cowan}, Preeti and {Croft}, Steve and {{\'C}uk}, Matija and {Daruich}, Felipe and {Daubard}, Guillaume and {Davenport}, James R.~A. and {Daylan}, Tansu and {Delgado}, Jennifer and {Devillepoix}, Hadrien A.~R. and {Doherty}, Peter E. and {Donaldson}, Abbie and {Drass}, Holger and {Deppe}, Stephanie JH and {Dubois-Felsmann}, Gregory P. and {Economou}, Frossie and {Eduardo}, Marielle R. and {Farnocchia}, Davide and {Frissell}, Maxwell K. and {Fedorets}, Grigori and {Fernandes}, Maryann Benny and {Fulle}, Marco and {Gerdes}, David W. and {Gibbs}, Alex R. and {Gillan}, A. Fraser and {Guy}, Leanne P. and {Hammergren}, Mark and {Hanushevsky}, Andrew and {Hernandez}, Fabio and {Hestroffer}, Daniel and {Hopkins}, Matthew J. and {Granvik}, Mikael and {Ieva}, Simone and {Irving}, David H. and {Jannuzi}, Buell T. and {Jimenez}, David and {Ramos Gomes-J{\'u}nior}, Altair and {Juramy}, Claire and {Kahn}, Steven M. and {Kannawadi}, Arun and {Kang}, Yijung and {Kryszczy{\'n}ska}, Agnieszka and {Kotov}, Ivan and {Koumjian}, Alec and {Krughoff}, K. Simon and {Lage}, Craig and {Lange}, Travis J. and {Levine}, W. Garrett and {Li}, Zhuofu and {Licandro}, Javier and {Lin}, Hsing Wen and {Lust}, Nate B. and {Lyttle}, Ryan R. and {Mahabal}, Ashish A. and {Mahlke}, Max and {Plazas Malag{\'o}n}, Andr{\'e}s A. and {Salazar Manzano}, Luis E. and {Marc}, Moniez and {Margoti}, Giuliano and {Mar{\v{c}}eta}, Du{\v{s}}an and {Menanteau}, Felipe and {Meyers}, Joshua and {Mills}, Dave and {Morato}, Naomi and {More}, Surhud and {Morrison}, Christopher B. and {Moulane}, Youssef and {Mu{\~n}oz-Guti{\'e}rrez}, Marco A. and {Newcomer F.}, M. and {O'Connor}, Paul and {Oldag}, Drew and {Oldroyd}, William J and {O'Mullane}, William and {Opitom}, Cyrielle and {Oszkiewicz}, Dagmara and {Page}, Gary L. and {Patterson}, Jack and {Payne}, Matthew J. and {Peloton}, Julien and {Pereira}, Chrystian Luciano and {Peterson}, John R. and {Polin}, Daniel and {Pollek}, Hannah Mary Margaret and {Polen}, Rebekah and {Qiu}, Yongqiang and {Ragozzine}, Darin and {Rajagopal}, Jayadev and {van Reeven}, vWouter and {Rice}, Malena and {Ridgway}, Stephen T. and {Rivkin}, Andrew S. and {Robinson}, James E. and {Ro{\.z}ek}, Agata and {Salnikov}, Andrei and {S{\'a}nchez}, Bruno O. and {Sarid}, Gal and {Schambeau}, Charles A. and {Scolnic}, Daniel and {Schindler}, Rafe H. and {Seaman}, Robert and {Jacques}, {\v{S}}ebag and {Shaw}, Richard A. and {Shugart}, Alysha and {Sick}, Jonathan and {Siraj}, Amir and {Sitarz}, Michael C. and {Sobhani}, Shahram and {Soldahl}, Christine and {Stalder}, Brian and {Stetzler}, Steven and {Swinbank}, John D. and {Szigeti}, L{\'a}szl{\'o} and {Tauraso}, Michael and {Thornton}, Adam and {Tonietti}, Luca and {Trilling}, David E. and {Trujillo}, Chadwick A.},
        title = "{NSF-DOE Vera C. Rubin Observatory Observations of Interstellar Comet 3I/ATLAS (C/2025 N1)}",
      journal = {arXiv e-prints},
         year = 2025,
        month = jul,
          eid = {arXiv:2507.13409},
        pages = {arXiv:2507.13409},
          doi = {10.48550/arXiv.2507.13409},
archivePrefix = {arXiv},
       eprint = {2507.13409},
 primaryClass = {astro-ph.EP},
       adsurl = {https://ui.adsabs.harvard.edu/abs/2025arXiv250713409C}
}

@ARTICLE{delaFuenteMarcos2025,
       author = {{de la Fuente Marcos}, R. and {Alarcon}, M.~R. and {Licandro}, J. and {Serra-Ricart}, M. and {de Le{\'o}n}, J. and {de la Fuente Marcos}, C. and {Lombardi}, G. and {Tejero}, A. and {Cabrera-Lavers}, A. and {Guerra Arencibia}, S. and {Ruiz Cejudo}, I.},
        title = "{Assessing interstellar comet 3I/ATLAS with the 10.4 m Gran Telescopio Canarias and the Two-meter Twin Telescope}",
      journal = {\aap},
         year = 2025,
        month = aug,
       volume = {700},
          eid = {L9},
        pages = {L9},
          doi = {10.1051/0004-6361/202556439},
archivePrefix = {arXiv},
       eprint = {2507.12922},
 primaryClass = {astro-ph.EP},
       adsurl = {https://ui.adsabs.harvard.edu/abs/2025A&A...700L...9D}
}

@ARTICLE{Cordiner_2025,
       author = {{Cordiner}, Martin A. and {Roth}, Nathaniel X. and {Kelley}, Michael S.~P. and {Bodewits}, Dennis and {Charnley}, Steven B. and {Drozdovskaya}, Maria N. and {Farnocchia}, Davide and {Micheli}, Marco and {Milam}, Stefanie N. and {Opitom}, Cyrielle and {Schwamb}, Megan E. and {Thomas}, Cristina A.},
        title = "{JWST detection of a carbon dioxide dominated gas coma surrounding interstellar object 3I/ATLAS}",
      journal = {arXiv e-prints},
         year = 2025,
        month = aug,
          eid = {arXiv:2508.18209},
        pages = {arXiv:2508.18209},
          doi = {10.48550/arXiv.2508.18209},
archivePrefix = {arXiv},
       eprint = {2508.18209},
 primaryClass = {astro-ph.EP},
       adsurl = {https://ui.adsabs.harvard.edu/abs/2025arXiv250818209C}
}

@ARTICLE{Rahatgaonkar_2025,
       author = {{Rahatgaonkar}, Rohan and {Carvajal}, Juan Pablo and {Puzia}, Thomas H. and {Luco}, Baltasar and {Jehin}, Emmanuel and {Hutsem{\'e}kers}, Damien and {Opitom}, Cyrielle and {Manfroid}, Jean and {Marsset}, Micha{\"e}l and {Yang}, Bin and {Buchanan}, Laura and {Fraser}, Wesley C. and {Forbes}, John and {Bannister}, Michele and {Bodewits}, Dennis and {Bolin}, Bryce T. and {Belyakov}, Matthew and {Knight}, Matthew M. and {Snodgrass}, Colin and {Bufanda}, Erica and {Dorsey}, Rosemary and {Ferellec}, L{\'e}a and {La Forgia}, Fiorangela and {Lippi}, Manuela and {Murphy}, Brian and {Nayak}, Prasanta K. and {Vander Donckt}, Mathieu},
        title = "{VLT observations of interstellar comet 3I/ATLAS II. From quiescence to glow: Dramatic rise of Ni I emission and incipient CN outgassing at large heliocentric distances}",
      journal = {arXiv e-prints},
         year = 2025,
        month = aug,
          eid = {arXiv:2508.18382},
        pages = {arXiv:2508.18382},
          doi = {10.48550/arXiv.2508.18382},
archivePrefix = {arXiv},
       eprint = {2508.18382},
 primaryClass = {astro-ph.SR},
       adsurl = {https://ui.adsabs.harvard.edu/abs/2025arXiv250818382R}
}

@ARTICLE{Jewitt2025,
       author = {{Jewitt}, David and {Hui}, Man-To and {Mutchler}, Max and {Kim}, Yoonyoung and {Agarwal}, Jessica},
        title = "{Hubble Space Telescope Observations of the Interstellar Interloper 3I/ATLAS}",
      journal = {\apjl},
         year = 2025,
        month = sep,
       volume = {990},
       number = {1},
          eid = {L2},
        pages = {L2},
          doi = {10.3847/2041-8213/adf8d8},
archivePrefix = {arXiv},
       eprint = {2508.02934},
 primaryClass = {astro-ph.EP},
       adsurl = {https://ui.adsabs.harvard.edu/abs/2025ApJ...990L...2J}
}

@article{Seligman2025,
	title        = {{Discovery and Preliminary Characterization of a Third Interstellar Object: 3I/ATLAS}},
	author       = {{Seligman}, Darryl Z. and {Micheli}, Marco and {Farnocchia}, Davide and {Denneau}, Larry and {Noonan}, John W. and {Hsieh}, Henry H. and {Santana-Ros}, Toni and {Tonry}, John and {Auchettl}, Katie and {Conversi}, Luca and {Devog{\`e}le}, Maxime and {Faggioli}, Laura and {Feinstein}, Adina D. and {Fenucci}, Marco and {Ferrais}, Marin and {Frincke}, Tessa and {Gillon}, Michael and {Hainaut}, Olivier R. and {Hart}, Kyle and {Hoffman}, Andrew and {Holt}, Carrie E. and {Hoogendam}, Willem B. and {Huber}, Mark E. and {Jehin}, Emmanuel and {Kareta}, Theodore and {Keane}, Jacqueline V. and {Kelley}, Michael S.~P. and {Lister}, Tim and {Mandt}, Kathleen and {Manfroid}, Jean and {Mar{\v{c}}eta}, Du{\v{s}}an and {Meech}, Karen J. and {Amine Miftah}, Mohamed and {Morgan}, Marvin and {Oca{\~n}a}, Francisco and {Pe{\~n}a-Asensio}, Eloy and {Shappee}, Benjamin J. and {Siverd}, Robert J. and {Taylor}, Aster G. and {Tucker}, Michael A. and {Wainscoat}, Richard and {Weryk}, Robert and {Wray}, James J. and {Yaginuma}, Atsuhiro and {Yang}, Bin and {Ye}, Quanzhi and {Zhang}, Qicheng},
	year         = 2025,
	month        = aug,
	journal      = {\apjl},
	volume       = 989,
	number       = 2,
	pages        = {L36},
	doi          = {10.3847/2041-8213/adf49a},
	eid          = {L36},
	archiveprefix = {arXiv},
	eprint       = {2507.02757},
	primaryclass = {astro-ph.EP},
	adsurl       = {https://ui.adsabs.harvard.edu/abs/2025ApJ...989L..36S}
}

@article{Do2018,
	title        = {{Interstellar Interlopers: Number Density and Origin of {\textquoteleft}Oumuamua-like Objects}},
	author       = {{Do}, Aaron and {Tucker}, Michael A. and {Tonry}, John},
	year         = 2018,
	month        = mar,
	journal      = {\apjl},
	volume       = 855,
	number       = 1,
	pages        = {L10},
	doi          = {10.3847/2041-8213/aaae67},
	eid          = {L10},
	archiveprefix = {arXiv},
	eprint       = {1801.02821},
	primaryclass = {astro-ph.EP},
	adsurl       = {https://ui.adsabs.harvard.edu/abs/2018ApJ...855L..10D}
}

@incollection{Fitzsimmons2024,
	title        = {{Interstellar Objects and Exocomets}},
	author       = {{Fitzsimmons}, Alan and {Meech}, Karen and {Matr{\`a}}, Luca and {Pfalzner}, Susanne},
	year         = 2024,
	booktitle    = {Comets III},
	pages        = {731--766},
	editor       = {{Meech}, Karen. J. and {Combi}, Michael. R. and {Bockel{\'e}e-Morvan}, Dominique and {Raymodn}, Sean. N. and {Zolensky}, Michael. E.},
	adsurl       = {https://ui.adsabs.harvard.edu/abs/2024come.book..731F}
}

@inproceedings{Lantz2004,
	title        = {{SNIFS: a wideband integral field spectrograph with microlens arrays}},
	author       = {{Lantz}, Blandine and {Aldering}, Greg and {Antilogus}, Pierre and {Bonnaud}, Christophe and {Capoani}, Lionel and {Castera}, Alain and {Copin}, Yannick and {Dubet}, Dominique and {Gangler}, Emmanuel and {Henault}, Francois and {Lemonnier}, Jean-Pierre and {Pain}, Reynald and {Pecontal}, Arlette and {Pecontal}, Emmanuel and {Smadja}, Gerard},
	year         = 2004,
	month        = feb,
	booktitle    = {Optical Design and Engineering},
	series       = {Society of Photo-Optical Instrumentation Engineers (SPIE) Conference Series},
	volume       = 5249,
	pages        = {146--155},
	doi          = {10.1117/12.512493},
	editor       = {{Mazuray}, Laurent and {Rogers}, Philip J. and {Wartmann}, Rolf},
	adsurl       = {https://ui.adsabs.harvard.edu/abs/2004SPIE.5249..146L}
}

@article{Jewitt2023ARAA,
	title        = {{The Interstellar Interlopers}},
	author       = {{Jewitt}, David and {Seligman}, Darryl Z.},
	year         = 2023,
	month        = aug,
	journal      = {\araa},
	volume       = 61,
	pages        = {197--236},
	doi          = {10.1146/annurev-astro-071221-054221},
	archiveprefix = {arXiv},
	eprint       = {2209.08182},
	primaryclass = {astro-ph.EP},
	adsurl       = {https://ui.adsabs.harvard.edu/abs/2023ARA&A..61..197J}
}
\bibliographystyle{aasjournalv7}



\end{document}